\documentclass[12pt,preprint]{aastex}

\usepackage{graphicx}
\usepackage{rotating}

\shorttitle{}
\shortauthors{Nesvorn'y et al.}

\begin{document}
\baselineskip 19.pt

\title{ExoMOD I. A Forward Model for the Orbital Architecture\\ of Kepler Multi-Planet Systems}

\author{David Nesvorn\'y$^{1}$, Daniel A. Yahalomi$^{2,3}$, David Kipping$^{4}$, Cristian Beaug\'e$^{5}$}

\affil{(1) Solar System Science \& Exploration Division, Southwest Research Institute, 1301 Walnut Street, 
  Suite 400,  Boulder, CO 80302, USA}

\affil{(2) MIT Kavli Institute for Astrophysics and Space Research, 70 Vassar Street, Cambridge, MA 02139, USA}

\affil{(3) Juan Carlos Torres Postdoctoral Fellow}

\affil{(4) Department of Astronomy, Columbia University, 550 W 120th Street, New York, NY 10027, USA}

\affil{(5) Instituto de Astronom\'{\i}a Te\'orica y Experimental (IATE), Observatorio Astron\'omico,
Universidad Nacional de C\'ordoba, Laprida 854, X5000BGR C\'ordoba, Argentina}

\begin{abstract}
  Transit observations only detect planets with favorable, near edge-on orientation of orbits as seen by a distant observer,
  which leaves much freedom for various interpretations in terms of the underlying planetary system architecture. Here we
  forward model transit observations of the Kepler telescope to characterize the orbital properties of close-in 
  planetary systems. We make sensible choices about the underlying distributions of planet radii, masses and orbital periods, 
  parameterize the orbital excitation with the Angular Momentum Deficit (AMD), and adopt an accurate method to account 
  for transit detection. The fits to Kepler's DR25 data are executed with {\tt MultiNest}. We find that the orbital period 
  distributions of Kepler singles and multis are statistically different from each other -- possibly a consequence of 
  Kepler's observational baseline. The observed gap complexity distribution is reproduced when planets in high 
  multiplicity systems ($m \geq 5$) are assigned ideally correlated period ratios. The high-multiplicity systems 
  probably retained a memory of their formation conditions. The gap complexity metric also helps to constrain the AMD 
  distribution. We find that systems with positive radius monotonicities typically harbor smaller planets, as 
  expected if the radius monotonicity is influenced by non-detections. A FGK dwarf in the Kepler field 
  should host $\simeq 2.4 \pm 0.2$ planets on average with radii $0.5 < R_{\rm pl}/R_\oplus < 7$ and orbital periods 
  $3<P_{\rm orb}<300$ d. For a detected planetary system, there is roughly a 50\% chance that Kepler transit observations
  missed at least one inner or intermediate-period planet. 
\end{abstract}

\section{Introduction}

Transit observations are blind to planets whose orbits are not nearly perpendicular to the sky plane. For multi-planet
systems, which are the main focus here, the detection probability depends on the mutual inclination of orbits. 
If orbits are perfectly aligned, the transit detection of an outer planet assures that all inner planets 
will be detected as well (assuming that they are large enough to produce a detectable photometric signal). 
Conversely, the detection probabilities of different planets in the same system are essentially unrelated if the 
mutual inclinations are very large. This dependence presents a fundamental uncertainty in interpreting 
transit observations: Are we not seeing more planets in a specific system because the mutual inclinations are large, 
or because these planets do not exist? Follow-up RV observations could address this issue but, unfortunately, many  
Kepler stars are too faint for precise Doppler measurements.

The degeneracy between the mutual inclination and planet multiplicity can be addressed, in the absence of additional information,
within the framework of a model where various system parameters are assumed to follow specific distribution functions.
For example, the Rayleigh distribution was often used for inclinations (e.g., Fabrycky \& Winn 2009, Lissauer et al. 2011, Fang
\& Margot 2012, Dawson et al. 2016, Zhu et al. 2018, Mulders et al. 2018), optionally adopting two different categories of
planets (e.g., the low- and high-inclination groups; He et al. 2019), or dynamics-informed criteria (e.g., the AMD-based
distribution; He et al. 2020). The intrinsic planet multiplicity was modeled with dichotomous, Poisson and Zipfian
distributions (Sandford et al. 2019 and the references therein).  All these choices are reasonable -- but they often lead to 
different results and interpretations in terms of the underlying inclination distribution 
(Zhu \& Dong 2021), and other system's parameters. The core of the problem is that the transit photometry alone does 
not uniquely constrain the inclination distribution of orbits and/or planet multiplicity (Tremaine \& Dong 2012).

The main idea of this work is to develop a population model of Kepler planets, hereafter referred to as ExoMOD, and use 
it to statistically predict Transit Timing Variations (TTVs) caused by the gravitational interaction of planets. The 
TTV algorithm and analysis will be described in Paper II (Nesvorn\'y et al., submitted). In Paper II, the population
model predictions will be compared to TTVs inferred from the Kepler transit data (Holczer et al. 2016, Hadden \& Lithwick 
2017, Ofir et al. 2018, Yahalomi et al. 2025) 
to address scientific questions related to the underlying multiplicity of planetary systems, mutual inclination 
distribution of planetary orbits, etc. Here we first describe four ExoMOD elements: the (1) input catalog of Kepler stars 
and planets (Section 2.1), (2) model description of the intrinsic properties of planetary systems (Section 2.2), 
(3) transit detection module (Section 2.3), and (4)  Bayesian inference for model selection and best-fit parameter 
estimation (Section 2.4). The results are discussed in Section 3. We estimate the planet occurrence rate in 
Section 4 and summarize the main findings of this work in Section 5. 

\section{Methods}

\subsection{Input catalog}

To define our Kepler dataset ${\cal D}_{\rm kep}$, we downloaded the Kepler DR25 Kepler Objects of Interest (KOI) 
catalog from the NASA Exoplanet Archive (Akeson et al. 2013, Christiansen et al. 2025).\footnote{Note that
the Kepler DR25 catalog does not incorporate the Gaia-based stellar parameters from Berger et al. (2020), nor 
the more recent update from Berger et al. (2026). It lists the stellar properties used by the original Kepler 
DR25 pipeline. This is not ideal, but it facilitates comparison with prior planet-occurrence studies that used 
Kepler DR25 as well (e.g., Mulders et al. 2018, He et al. 2019, 2020). We will update ExoMOD with the data from 
Berger et al. (2026) in the near future.} We only selected KOIs for 
which the ``Disposition using Kepler Data'' was not reported as FALSE POSITIVE. To focus on FGK dwarfs, we required that the 
stellar surface gravity $g$ and stellar radius $R_*$ satisfied $\log g > 4$ (in cgs) and $0.7 < R_*/R_\odot < 1.6$, 
respectively, where $R_\odot$ is the solar radius. Additionally, we only included planetary candidates with the 
estimated physical radius $0.5<R/R_\oplus<7$, where $R_\oplus$ is the Earth's radius, and the orbital period $3<P_{\rm orb}<300$ d. 
For comparison, Mulders et al. (2018) used $0.5<R/R_\oplus<6$ and $2<P_{\rm orb}<400$ d and He et al. (2019, 2020, 2021) 
used $0.5<R/R_\oplus<10$ and $3<P_{\rm orb}<300$~d. At least three observed transits were required for a planet to be 
included in ${\cal D}_{\rm kep}$. 

The selection described above yielded $n({\cal D}_{\rm kep})=2761$ planets in total most of which are dispositioned 
as CONFIRMED. Hereafter, ${\cal D}_{{\rm kep},m}$ denotes the set of Kepler planets observed in systems of multiplicity 
$m$. We have $n({\cal D}_{{\rm kep},1})=1591$ singles (observed multiplicity $m=1$), $n({\cal D}_{{\rm kep},2})=657$, 
$n({\cal D}_{{\rm kep},3})=327$, $n({\cal D}_{{\rm kep},4})=129$, $n({\cal D}_{{\rm kep},5})=45$, $n({\cal D}_{{\rm kep},6})=12$,
and $n({\cal D}_{\rm kep,m})=0$ for $m \geq 7$. For each of these planets, we recorded: the (1) observed multiplicity 
of systems to which they belong, (2) orbital period $P_{\rm orb}$, (3) inferred planet radius $R_{\rm pl}$, (4) transit 
depth $\delta$, (5) transit duration $T_{\rm dur}$, (6) signal-to-noise ratio (SNR) of detection. For planets in the 
same system with observed multiplicity $m \geq 2$, we also computed: the (7) orbital period ratio of neighbor planets 
$f_P=P_j/P_{j-1}$, and (8) planet radius ratio $f_R=R_j/R_{j-1}$, where $j$ and $j-1$ stand for the outer and inner 
planets, respectively.

\subsection{Forward model}

Here we define our intrinsic model ${\cal M}_{\rm int}$. This involves a number of semi-arbitrary choices because 
the statistical description of underlying (intrinsic) planet architectures is subject to a considerable uncertainty 
(Tremaine \& Dong 2012). Planet formation simulations could be used to inform expected planet properties but the 
existing models involve unknown physical parameters, e.g. related to the planet migration in protoplanetary 
disks, and do not have the required predictive power (e.g., Izidoro et al. 2021). In this work, we adopt several 
different models and employ the Bayesian model selection framework (Section 2.4) to determine which them provides 
a better representation of the Kepler data. 

\subsubsection{Stellar parameters}

Following Sandford et al. (2019), we queried Kepler targets for stars that match the filters imposed above (Mathur et al. 
2017, Berger et al. 2018), obtaining 107,983 FGK dwarfs. This catalog is used to generate the stellar parameters in ExoMOD. 
In this work, we only study the planet occurrence around the combined population of Kepler's FGK dwarfs without making 
any distinction between F, G and K hosts (cf. He et al. 2021).
To set up a planetary system in ExoMOD, we first select a random star from the filtered 
stellar catalog described above. This gives us the stellar radius $R_*$, stellar mass $M_*$, and a log of Kepler quarters 
in which the star was observed. These parameters are used to determine the transit detectability for each model 
planet (Section 2.3).

\subsubsection{Multiplicity}

A planetary system of $m$ planets is injected around each target star. To make a distinction between the intrinsic 
(planet) multiplicity and observed (tranet) multiplicity, we denote the set of planets in systems with the intrinsic 
multiplicity $m_{\rm int}$ as ${\cal M}_{{\rm int},m}$,\footnote{Note the difference between ${\cal M}_{{\rm int},m}$ and 
$m_{\rm int}$. The intrinsic multiplicity $m_{\rm int}$ is just an integer denoting the number of planets in a system, 
whereas ${\cal M}_{{\rm int},m}$ is the set of planets, with all their properties, in systems of multiplicity 
$m_{\rm int}$. We specify the index $m$ in ${\cal M}_{{\rm int},m}$ because the planet properties generated by our method 
depend on their parent system multiplicity.} and the set of planets in systems with the observed multiplicity 
$m_{\rm obs}$ as ${\cal M}_{{\rm obs},m}$. The properties of ${\cal M}_{{\rm int},m}$ are defined in the following 
sections. ${\cal M}_{{\rm obs},m}$ is obtained from ${\cal M}_{{\rm int},m}$ by accounting for transit detection in 
Section 2.3. 

We also need to make a distinction between the {\it system} multiplicity, $m^{\rm sys}$, a system's characteristic 
that defines the number of planets in the system, and the {\it planet} multiplicity, $m^{\rm pla}$, defined here as the 
property of a planet indicating the number of planets in the planet's parent system. The two distributions are 
related via ${\rm Pr}(m^{\rm pla}) \propto m^{\rm sys} {\rm Pr}(m^{\rm sys})$. We omit superscripts ``sys'' and ``pla''
in the rest of the paper, because we always use the planet multiplicity $m^{\rm pla}$. Any $m$ appearing hereafter 
therefore stands for $m^{\rm pla}$.

Two different models were adopted for $m_{\rm int}$ (we drop the subscript specifier below as it is understood that 
the {\it intrinsic system} multiplicity distribution is defined here).  
Our first choice is a truncated Poisson distribution, where the probability of forming $m$ planets in the 
same system is 
\begin{equation}
{\rm Pr}(m|\lambda_{\rm pl}) = { \lambda_{\rm pl}^m \exp(-\lambda_{\rm pl}) \over m! }\ {\rm for}\ 0 \leq m \leq m_{\rm max}   
\label{poisson}
\end{equation}  
and 0 otherwise, with $m_{\rm max}=10$. Here, $\lambda_{\rm pl}$ is the mean number of planets per system in the 
(non-truncated) Poisson distribution. With the truncation, the mean number of planets per system differs 
from $\lambda_{\rm pl}$. The Poisson distribution favors systems with $m \sim \lambda_{\rm pl}$, but we do not know a 
priori whether this is a good model for the real multiplicity distribution.

As we are interested in the overall planet occurrence rate, we also generate stars with no planets ($m_{\rm int}=0$). 
This allows us, at least in principle, to determine the overall occurrence rate of planets. It has to be 
emphasized, however, that this inference sensitively depends on the choice of the planet multiplicity model 
and its implications for the number of stars with no planets. 

Our second choice is the Zipfian model with a discrete power-law probability distribution
\begin{equation}
{\rm Pr}(m|\beta_{\rm zipf}) \propto  m^{\beta_{\rm zipf}}\ {\rm for}\  1 \leq m \leq m_{\rm max}   
\label{zipfian}
\end{equation} 
and 0 otherwise. In this case, all stars have at least one planet (note that ${\rm Pr}(m|\beta_{\rm zipf})$ is 
undefined for $\beta_{\rm zipf}<0$ and $m=0$). Sandford et al. (2019) found 
$\beta_{\rm zipf} = -1.86 \pm 0.3$ strongly favoring systems with low multiplicities. They pointed out that a 
dichotomous distribution (e.g., Ballard \& Johnson 2016) is not needed to explain Kepler observations. 
Other models investigated in Sandford et al. (2019) are somewhat less relevant (e.g., the constant model is probably 
too simplistic, the uniform model is a special case of the Zipfian distribution with $\beta_{\rm zipf}=0$, etc.).
They are not investigated here. 

\subsubsection{Planetary radii}

The physical radius of the innermost planet, $0.5 \leq R_{1} \leq 7$ $R_\oplus$, follows a broken power-law 
\begin{equation}
{\rm Pr}(R_1) \propto R_1^{-\alpha_{\rm R}}\ {\rm for}\ R_1 \leq R_{\rm break}   
\label{radius1}
\end{equation} 
and 
\begin{equation}
{\rm Pr}(R_1) \propto R_1^{-\beta_{\rm R}}\ {\rm for}\ R_1 > R_{\rm break} \ ,   
\label{radius2}
\end{equation} 
where $R_{\rm break}$, $\alpha_{\rm R}$ and $\beta_{\rm R}$ are model parameters (Youdin 2011, Howard et al. 2012, 
Petigura et al. 2013, Mulders et al. 2018). 


To respect the `peas-in-a-pod' covariance (Millholland et al. 2017, Weiss et al. 2018, Murchikova \& Tremaine
2020, Weiss et al. 2022, Otegi et al. 2022, Goyal \& Wang 2022), 
the radii of outer planets $R_j$, $2 \leq j \leq m$, are chosen 
from a normal distribution with the mean $R_1$ and standard deviation $\sigma_R$:
\begin{equation}
{\rm Pr}(R_j|\sigma_R) = {1 \over \sqrt{2 \pi} \sigma_R} \exp \left(- { (R_j - R_1)^2 \over 2 \sigma_R^2} \right )     \ .
\label{peas} 
\end{equation} 
The parameter $\sigma_R$ controls the spread of planetary radii as a fraction of $R_1$; it can be large, thereby 
accounting for the possibility of no correlation (e.g., Zhu 2019). A limitation of this approach is that it ignores  
any size ordering of planets with orbital radius. The radius monotonicity is discussed below and in Section 3. 

Whereas the normal distribution in Eq. (\ref{peas}) respects the radius covariance, we noticed that 
the tails of the normal distribution are not heavy enough to fully reproduce observations. Thus, 
as a slight improvement over Eq. (\ref{peas}), we adopt Student's $t$-distribution
\begin{equation}
{\rm Pr}(R_j|\nu_R) \propto \left ( 1 + { (R_j-R_1)^2 \over \nu_R } \right )^{-{\nu_R +1 \over 2}}   
\label{peas2} 
\end{equation} 
with $\nu_R=0.5$, rescaled to the variance $\sigma_R$.\footnote{We also tested the clustered log-normal 
distribution from He et al. (2020) but did not find acceptable fits with this distribution to the 
planet radius ratios. The log-normal distribution gives the best-fit radius ratio distribution that is 
not as peaked as the observed one. A similar problem can be seen in Fig. 2 of He et al. (2020) for the 
transit depth ratio distributions.}     

The double-sided power-law radius distribution in Eqs. (\ref{radius1}) and (\ref{radius2})
is designed to capture the turnover in planet occurrence 
seen at around mini-Neptune radii, as reported in numerous studies (Fressin et al. 2013, Petigura et al. 2013, 
Foreman-Mackey et al. 2014). It does not account for the radius valley of Fulton et al. (2017). 
This effect was only revealed by substantial improvements to the precision of measured stellar radii, and 
should not be influential enough to significantly affect our study (He \& Ford 2026). 
The planet radius distributions in single 
and multi-planet systems are nearly identical (e.g., Lissauer et al. 2024); we therefore disregard any potential 
dependence of ${\rm Pr}(R)$ on $m_{\rm int}$.  

Several studies have found that exoplanets are preferentially arranged with larger planets exterior to smaller
planets (Ciardi et al. 2013, Millholland et al. 2017, Kipping 2018, Weiss et al. 2018). We tested models with 
strict radius-period ordering ($R_j/R_{j-1}>1$) but found that these models are disfavored because they produced 
a strong preference toward $R_j/R_{j-1}>1$ of the detected planets, too strong when compared with the Kepler data.
We therefore decided, for the sake of simplicity, to ignore any correlation of planet size with orbital period
in ${\cal M}_{\rm int}$. The radius-period ordering arises in ${\cal M}_{\rm obs}$ as more distant planets have fewer 
transits, lower SNR, and tend be larger to be detected. This effect contributes the observed radius-period 
monotonicity but may not fully explain it (Section 3).

\subsubsection{Planetary masses}

The planetary masses, $M_j$, as needed for the AMD (Section 2.2.6) and TTV analysis (Paper II), 
are computed from the planetary radii following  
\begin{equation}
M_j= 0.972 R_j^{3.58}\ {\rm for}\ R_j < 1.23\,R_\oplus  
\end{equation}
and 
\begin{equation}
M_j= 1.437 R_j^{1.70}\ {\rm for}\ R_j>1.23\,R_\oplus\
\label{chen}   
\end{equation}
(Chen \& Kipping 2017). Eq. (\ref{chen}) would give excessive masses for very large planets but this
is irrelevant here as we have $R_j<7\,R_\oplus$ (Section 2.2.3).  We also tested other prescriptions (Wolfgang 
et al. 2016, Ning et al. 2018, Otegi et al. 2020, Muller et al. 2024) and found that this did not significantly 
influence the results.

\subsubsection{Orbital periods}

The orbital period of the innermost planet, $3\ {\rm d} < P_1 < 300$ d, is drawn from a broken power-law 
distribution
\begin{equation}
{\rm Pr}(P_1) \propto P_1^{-\alpha_P}\ {\rm for}\ P_1 \leq P_{\rm break}   
\end{equation} 
and 
\begin{equation}
{\rm Pr}(P_1) \propto P_1^{-\beta_P}\ {\rm for}\ P_1 > P_{\rm break}\ ,   
\end{equation} 
where $P_{\rm break}$, $\alpha_P$ and $\beta_P$ are model parameters (Mulders et al. 2018). The break reflects the 
observed flattening of sub-Neptune occurrence rates around an orbital period of ten days (Howard et al. 2012,
Mulders et al. 2015), possibly due the inner disk edge or planet trap (e.g., Millan-Gabet et al. 2007). 
 
The Hill radius of neighbor planets $j$ and $j-1$ is defined as
\begin{equation}
R_{\rm H}= \left( {M_j + M_{\rm j-1} \over 3 M_* } \right)^{1/3}  a_{j-1} \ .   
\end{equation} 
The planet $j \geq 2$ is placed at some multiple of Hill radii, $f_{\rm H}$, outside of planet $j-1$, 
\begin{equation}
a_j = a_{j-1} + f_{\rm H} R_{\rm H}\ . 
\label{aj}
\end{equation}
Here, $f_{\rm H}$ is assumed to follow the normal distribution
\begin{equation} 
{\rm Pr}(f_{\rm H}) = {1 \over \sqrt{2 \pi} \Sigma_{\rm H} } \exp \left 
(- {(f_{\rm H} - \mu_{\rm H})^2 \over 2 \Sigma_{\rm H}^2} \right ) 
\end{equation}
with two parameters $\mu_{\rm H}$ and $\Sigma_{\rm H}$. In Section 3, we prefer to report the normalized variance 
$\sigma_{\rm H} = \Sigma_{\rm H} / \mu_{\rm H}$.
The method described here is designed to produce the approximately log-normal orbital period distribution 
(Malhotra 2015) and orbital period ratios of Kepler planets. For comparison, He et al. (2019, 2020, 2021) 
distributed {\it all} planets in the same system using the log-normal distribution.

\subsubsection{The AMD model}

We employ the AMD model of He et al. (2020) to assign orbital eccentricities and mutual inclinations to planetary
orbits. This works as follows. When the planet $2\leq j \leq m_{\rm int}$ is placed on outside of the planet $j-1$ following the method 
described in the previous section, we first check whether the separation of the two planets is large enough for 
them to be stable if they had circular orbits (Wisdom 1980, Deck et al. 2013). Defining $\mu_j=M_j/M_*$ and 
$\alpha_j=a_{j-1}/a_j<1$, the stability criterion is 
\begin{equation}
\alpha_j<\alpha_{\rm crit} = 1-1.46(\mu_{j-1}+\mu_j)^{2/7} \ .
\end{equation}
If the criterion fails, the outer planet $j$ is rejected and we proceed by generating new $a_j$ from
Eqs. (\ref{aj}) and (13). If the stability criterion is satisfied, we move to the next planet until the last planet 
$j=m_{\rm int}$ is accepted. 

We define the critical AMD for the system of $m_{\rm int}$ planets to be stable according to Laskar 
\& Petit (2017) and Petit et al. (2017).\footnote{The AMD model is applied here to close-in planets; we 
disregard any large planets ($R/R_\oplus>7$) at large orbital radii ($P_{\rm orb}>300$ d). The terrestrial planets in 
the solar system are AMD-unstable, when the AMD is computed over all eight planets, because the outer planets represent a 
large AMD buffer. In this sense, the AMD model adopted here is is only approximate.} This involves two 
conditions. The first one assures that the AMD 
in the whole system is low enough such that if all AMD is transferred a pair of planets, the two orbits remain 
non-crossing and a collision between the planets cannot happen (Laskar \& Petit 2017). The second one 
guarantees  that all pairs of planetary orbits are stable against the overlap of mean motion resonances 
(MMRs; Petit et al. 2017). Mathematical expressions for these two criteria were given in Section 2.3.1 of 
He et al. (2020) and we do not repeat them here. The critical AMD value,  AMD$_{\rm crit}$, defines the 
maximum amount of AMD for a given set of planetary masses and semimajor axes such that the orbits are 
stable.         

Based on AMD$_{\rm crit}$, we assign the total AMD to a planetary system. In our base models, we set
\begin{equation}
{\rm AMD}_{\rm tot} = f_{\rm AMD} {\rm AMD}_{\rm crit}\ ,
\end{equation}
where $f_{\rm AMD}=0.3$ or 1 is a fixed AMD fraction. He et al. (2020) adopted $f_{\rm AMD}=1$ in their fiducial 
model,\footnote{The critical AMD model with $f_{\rm AMD}=1$ could apply if a planetary system starts 
with relatively high AMD, planets evolve onto crossing orbits and undergo collisions, and eventually 
reach orbital stability near ${\rm AMD}_{\rm crit}$. In reality, however, if planets migrated 
on nearly circular and nearly coplanar orbits into compact resonant chains (Izidoro et al. 2017, 2021; Goldberg 
\& Batygin 2022), the emerging AMD would have been small.} but 
they also tested $f_{\rm AMD}<1$. 
Additional options are investigated in our auxiliary models in Section 3.2, where instead of fixing $f_{\rm AMD}$ 
we let {\tt MultiNest} to fit for it. We find that this produces meaningful results only if the gap complexity
(Gilbert \& Fabrycky 2020) is used, in addition to all other metrics, to constrain the model.



We proceed by distributing ${\rm AMD}_{\rm tot}$ among individual planets. Following the principle of
equipartition, each planet $1 \leq j \leq m_{\rm int}$ in a system of multiplicity $m_{\rm int}$ is given the same share, 
${\rm AMD}_j = {\rm AMD}_{\rm tot}/m_{\rm int}$. Consequently, more massive planets have less excited orbits 
in the model, as often seen in dynamical simulations (e.g., Izidoro et al. 2017, 2021; Hansen \& Murray 2012, 2013). 
The orbital eccentricities $e_j$ and mutual inclinations $i_j$ of each planet are computed from ${\rm AMD}_j$ 
following the method described in Section 2.3.2. of He et al. (2020; their Eq. (33)). For $m_{\rm int}=1$, we 
define $i_1=0$ and use the Rayleigh distribution with the parameter $\sigma_e$ for $e_1$ (He et al. 2020). The 
mean eccentricity is $\langle e_1 \rangle = \sigma_e \sqrt{\pi/2}$ in this case. The results were found 
insensitive to $\sigma_e$.

The AMD model provides a convenient way to assign $e_j$ and $i_j$ to planetary orbits such that they are 
stable by design. In addition, prior analyses of normalized transit durations hint on decreasing orbital 
eccentricities in planetary systems with higher multiplicities (Steffen et al. 2010, Fang \& Margot 2012,
Fabrycky et al. 2014, He et al. 2019, 2020). The inverse correlation between eccentricities and 
multiplicity were also inferred from the analysis of radial velocity data (Limbach \& Turner 2015,
Zinzi \& Turrini 2017). This trend qualitatively arises in the AMD model as the total AMD is split among 
$m_{\rm int}$ orbits; planets in the higher multiplicity systems then have lower individual AMDs and lower 
$e_j$ values. 


\subsection{Transit detection}

The transit method is inherently biased due to both geometric and detection driven biases. To account for 
the geometric detection, we randomly select distant observer's location on the celestial sphere and compute
the impact parameter $b_j$ for each planet. Planets with $b_j<1$ are deemed to satisfy the geometric condition 
of transit. At least three transits are required for a planet to be included in ${\cal M}_{\rm obs}$ (the same 
criterion was adopted to define our Kepler dataset ${\cal D}_{\rm kep}$; Section 2.1). For the transiting planets, 
the transit duration $T_{\rm dur}$ is computed from the analytic formulas given in Millholland et al. (2021). 

The work of Kipping \& Sandford (2016) is used to determine the SNR of transit lightcurve as a function of 
basic transit parameters. This method accurately accounts for grazing transits. The final expressions for 
the SNR of a single transit are given in Eq. (17) and (18) of  Kipping \& Sandford (2016).  The SNR scales with the square 
root of the number of observed transits. The probability of planet detection is derived from the SNR using 
updated parameters from the Fressin et al. (2013) ramp.\footnote{The transit detection probability is 
reduced in multiplanet systems (Zink et al. 2019), but that is not something we investigate here.}
The transit detection algorithm is applied to ${\cal M}_{\rm int}$, producing the set of observed model planets
${\cal M}_{\rm obs}$, which can be compared to ${\cal D}_{\rm kep}$.

\subsection{Bayesian inference with {\tt MultiNest}}

The {\tt MultiNest} algorithm (Feroz \& Hobson 2008, Feroz et al. 2009) is used to execute the model 
optimization, parameter estimation and model selection. {\tt MultiNest} is a Bayesian inference tool that 
calculates the evidence and produces posterior samples from distributions in high dimensions. {\tt MultiNest} 
implements a multimodal nested sampling routine Skilling et al. (2014) designed to compute the Bayesian 
evidence in complex parameter space in an efficient manner. The algorithm significantly outperforms the 
existing Markov Chain Monte Carlo techniques, which generally have trouble accurately estimating the 
evidence term. The model selection is performed by comparing the Bayesian evidence of each model.
See below for the definition of the likelihood term. 
 
In our base models, the summary statistics include: the (1) total number of detected planets $N_{\rm p}$ as a 
fraction of the number of target stars $N_*$, $f=N_{\rm p}/N_*$, (2) observed multiplicity distribution, 
$n({\cal M}_{{\rm obs},m})$, where ${\cal M}_{{\rm obs},m}$ is the set of detected planets in systems of observed 
multiplicity $m_{\rm obs}$, (3) observed orbital period distribution, ${P_{\rm orb}}$, and (4) observed orbital period ratio 
distribution, ${P_j/P_{j-1}}$. More complex models also include constraints from the transit depth, transit 
depth ratio, transit duration, and/or period-normalized transit duration distributions.  Additional 
metrics were adapted from Gilbert \& Fabrycky (2020) (Section~3).

Given the focus of this work, which includes constraints from TTVs (Paper II), we give a preference to 
simple models that are optimized on a relatively small number of metric parameters (e.g., the four metrics highlighted 
above). Once these simple models are obtained, however, they are shown to be consistent with a larger ensemble
of metrics as well. This is primarily done to show the predictive power of simple models and to demonstrate
that the choices described in Section 2.2 are reasonable. 

We use a binning approach to defining the likelihood term. In our base models, the likelihood term is 
defined in 3D space of normalized multiplicity, orbital period, and orbital period ratio. Here, 
$n({\cal M}_{{\rm obs},m})$ is normalized from the number of model stars -- we typically use $10^7$ stars 
in a single trial to reduce the statistical noise -- to the number of Kepler stars (107,983 in total; 
Section 2.2.1). This effectively folds the metric $f=N_{\rm p}/N_*$ in the normalized multiplicity vector. 
The 3D space is divided into bins, the number of Kepler planets $n_k$ in each bin $k$ is obtained from 
${\cal D}_{\rm kep}$, and the expected number of model planets $\lambda_k$ in bin $k$ is computed 
from ${\cal M}_{\rm obs}$. 

Assuming the Poisson statistics, the probability of drawing $n_k$ objects is 
\begin{equation}
p_k(n_k) = {\lambda_k^{n_k} \exp(-\lambda_k) \over n_k!} \ . 
\end{equation}    
The joint probability over all bins is then 
\begin{equation}
P = \prod_k {\lambda_k^{n_k} \exp(-\lambda_k) \over n_k!} \ . 
\end{equation}   
The log-likelihood can therefore be defined as 
\begin{equation}
{\cal L} = \ln P = - \sum_k \lambda_k + \sum_k n_k \ln \lambda_k \ ,
\label{like}
\end{equation}
where we dropped the constant term $\sum_k \ln (n_k!)$.  

Ideally, we would like to use a full 3D grid such that any correlation between
different metrics can properly be represented (e.g., the orbital period distribution of Kepler planets 
depends on the observed multiplicity; Lissauer et al. 2024). This turns out to be impractical. 
In a typical model trial with $10^7$ model stars, roughly 200,000-300,000 planets are detected. 
Thus, for example, if there are 1,000 bins in total, 10 grid points in each dimension, we only expect 
200-300 planets in a bin on average, which sets $\lambda_k$. If $\lambda_k$ is that small, however, there 
are relatively strong statistical fluctuations and ${\cal L}$ significantly changes in random trials 
even if the planet system parameters are held fixed. We would need to increase the number of 
model stars but that would imply excessive CPU requirements. Clearly, some sort of smoothing is needed.

In prior works, separate distance functions based on the Kolmogorov-Smirnov and Anderson-Darling 
tests were used for each marginal distribution (e.g., Mulders et al. 2018, He et al. 2019, 2020; 
Sandford et al. 2019). These individual distance terms were  
combined -- as a linear weighted sum -- into a single distance function, which was then minimized with an
optimization algorithm. The advantage of this approach is that the marginalized 
distributions are less noisy, which reduces the stochasticity in the distance evaluation.
The disadvantage is that the marginalized distributions cannot capture any correlations between different 
metrics. 

As a compromise, here we project 3D metrics space into two 2D spaces, $({n({\cal M}_{{\rm obs},m}),P_{\rm orb}})$
and $({n({\cal M}_{{\rm obs},m\geq2}),P_j/P_{j-1}})$, and consider bins in each space separately.\footnote{The 
method can be easily extended to $N$ dimensions where we would have $N-1$ 2D grids.} This reduces
the number of grid points, increases $\lambda_k$, and reduces the model noise. With this setup, we are able 
to capture any dependence on planet multiplicity. Even with this method, however, the model is still
grainy, often preventing {\tt MultiNest} from converging. We therefore use {\tt MultiNest} to sample the prior space 
and record all trial evaluations of ${\cal L}$. In the second pass, {\tt MultiNest} reads this evaluations 
and computes ${\cal L}$ for any new trial by smoothing over the first-pass 
trials.\footnote{The smoothing is achieved
by averaging the first-pass trial ${\cal L}$ values in the hyperspherical neighborhood of the second-pass 
trial. First, {\tt MultiNest} renormalizes the actual physical parameter space onto a unit hypercube, where 
every parameter varies from 0 to 1. The optimal hypersphere radius, $R_{\rm hypsph}$, in 
normalized space is then established by testing: not too small for effective smoothing, not too big to 
avoid artifacts. We generate a large enough number of trial evaluations in the first pass ($\sim 3\times10^7$) 
such that $R_{\rm hypsph}=0.01$-0.05 can be used, which is deemed to be satisfactory.}            
  
\section{Results}

\subsection{Base models}

Our base models adopt different multiplicity distributions (Poisson or Zipfian) and AMD fractions 
($f_{\rm AMD}$). There are four base models in total: ${\cal M}_{219}$ with the Poisson distribution
and $f_{\rm AMD}=1$, ${\cal M}_{220}$ with the Poisson distribution and $f_{\rm AMD}=0.3$, ${\cal M}_{221}$ with 
the Zipfian distribution and $f_{\rm AMD}=1$, and ${\cal M}_{222}$ with the Zipfian distribution and 
$f_{\rm AMD}=0.3$ (Table 1).\footnote{We tested over two hundred models in this project with the 
three-digit index number uniquely identifying each model. This is the notation used here as well.}     
The models with $f_{\rm AMD}=1$ are similar to the maximum AMD model investigated 
in He et al. (2020). We set $f_{\rm AMD}=0.3$ in ${\cal M}_{220}$ and ${\cal M}_{222}$ to test low-AMD models
as well, because the AMD distribution of Kepler planets is not known a priori.
There are six priors: $\lambda_{\rm pl}$, $f_{\rm H}$, $\sigma_{\rm H}$, $P_{\rm break}$, $\alpha_P$, $\beta_P$.
The parameters defining the planet radius distribution are held fixed (Section 2.2.3; Table 2).
Following the procedure outlined in Section 2.4, {\tt MultiNest} evaluates the likelihood term in 
$(n({\cal M}_{{\rm obs},m}), P_{\rm orb})$ and $(n({\cal M}_{{\rm obs},m}),P_j/P_{j-1})$.\footnote{We used 
$N_{\rm grid}=10$ grid points in $n({\cal M}_{{\rm obs},m})$, $1\leq m \leq 10$, $N_{\rm grid}=20$ in $P_{\rm orb}$ 
and $N_{\rm grid}=36$ in $P_j/P_{j-1}$. The larger number of grid points in $P_j/P_{j-1}$ was helpful to 
adequately resolve the distribution for $P_j/P_{j-1}<2$. We also tested a larger number of grid 
points and found that the results were comparable, except that the statistical noise increased.} 

The Bayes factors are similar for our four base models (Table 1); we therefore cannot rule out any of them. 
In the following text, we highlight ${\cal M}_{219}$ for reference mainly because this model is the most
similar to the maximum AMD model investigated in He et al. (2020). The other base models are discussed
only when they present some notable differences relative to ${\cal M}_{219}$.

\begin{figure}
\epsscale{0.8}
\plotone{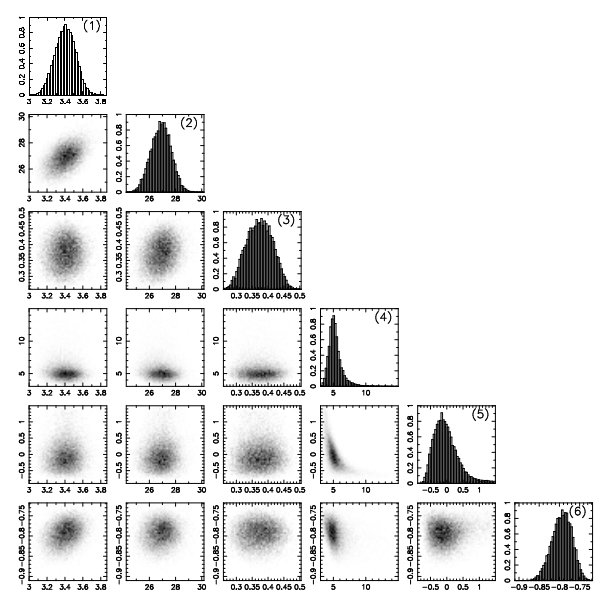}
\caption{A corner plot of posteriors from the ${\cal M}_{219}$ model. See Table 2 for the definition 
of parameters (1) to (6), their prior bounds, and the best-fit values obtained from {\tt MultiNest}.}
\label{corner1}
\end{figure}

Figure \ref{corner1} shows the corner plot and Table 2 lists the best-fit parameters from ${\cal M}_{219}$.
The ${\cal M}_{219}$ posteriors are well behaved. We identify an inverse correlation between $P_{\rm break}$ 
and $\alpha_P$ (parameters (4) and (5) in Fig. \ref{corner1}) with $P_{\rm break} \rightarrow 10$ d 
implying $\alpha_P  \rightarrow -0.6$ (i.e., a transition at larger orbital radii requires a steeper 
slope below the transition). The preferred values are $P_{\rm break} \simeq 5.2$ d and $\alpha_{\rm P} 
\simeq 0$ (a flat distribution below $P_{\rm break}$). No other correlations are apparent in Fig. 
\ref{corner1}.

\begin{figure}
\epsscale{0.8}
\plotone{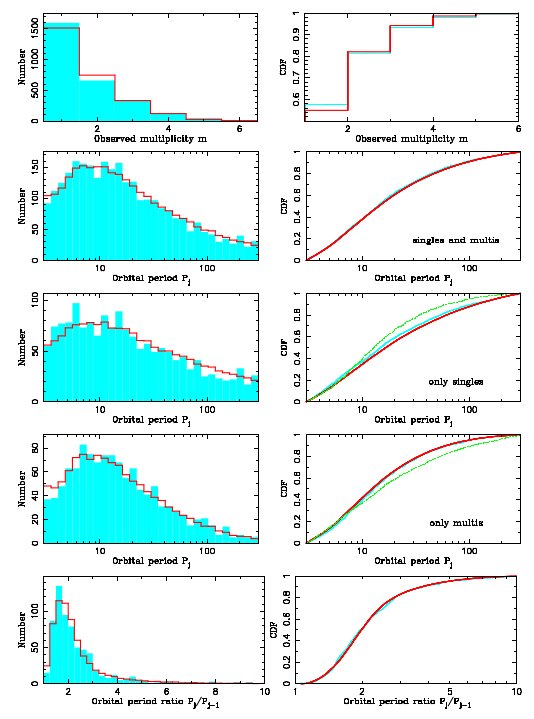}
\caption{Marginalized best-fit ${\cal M}_{\rm obs}$ from the ${\cal M}_{219}$ model (red lines) is
compared to marginalized ${\cal D}_{\rm kep}$ (blue histograms and blue lines). The two columns of plots
show the differential (left) and cumulative distribution functions (right). From top to bottom,
the rows of plots are the: (1) observed multiplicity ($m_{\rm obs}$), (2) orbital periods of all planets
($P_{\rm orb}$), (3) orbital periods of singles (planets in systems of observed multiplicity $m_{\rm obs}=1$), 
(4) orbital periods of multis (observed multiplicity $m_{\rm obs} \geq 2$), and (5) orbital period ratios 
of neighbor planets ($P_j/P_{j-1}$ for $2 \leq j \leq m_{\rm obs}$). The green dashed lines in rows (3) and 
(4) are plotted for reference: the period distribution of multis in (3) and singles in (4). Note 
that the top plots show the planet multiplicity, as defined in Section 2.2.2, and not the system
multiplicity as in Zhu et al. (2018) or He et al. (2020).}
\label{model219a}
\end{figure}

\begin{figure}
\epsscale{0.8}
\plotone{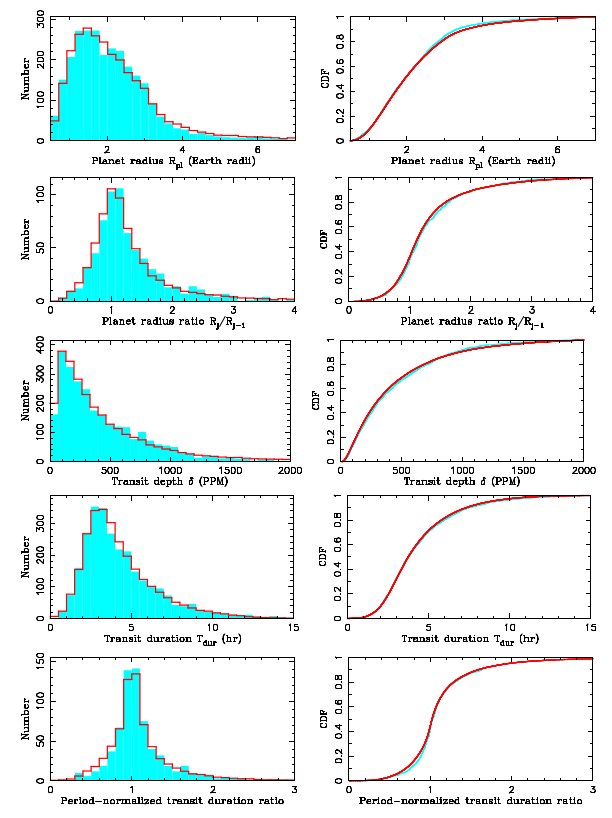}
\caption{Marginalized best-fit ${\cal M}_{\rm obs}$ from the ${\cal M}_{219}$ model (red lines) is
compared to marginalized ${\cal D}_{\rm kep}$ (blue histograms and blue lines). The two columns of plots
show the differential (left) and cumulative distribution functions (right). From top to bottom,
the rows of plots show: the (1) planet radius ($R_{\rm pl}$), (2) radius ratio of neighbor planets 
($R_j/R_{j-1}$ for $2 \leq j \leq m_{\rm obs}$), (3) transit depth ($\delta$), (4) transit duration 
($T_{\rm dur}$), and (5) period-normalized transit duration ratio ($\xi$).}
\label{model219b}
\end{figure}

\begin{figure}
\epsscale{0.8}
\plotone{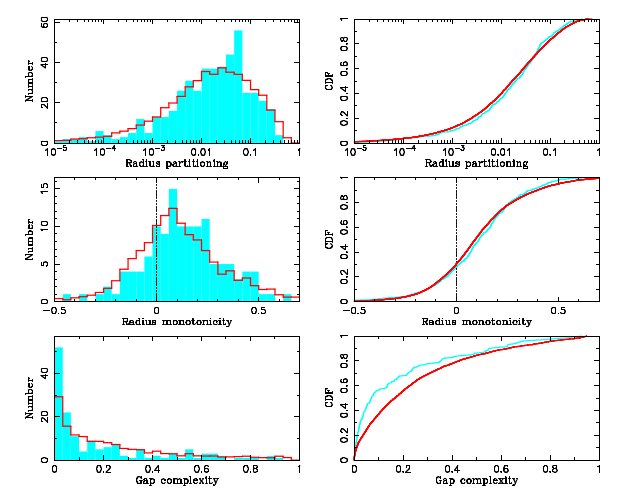}
\caption{Marginalized best-fit ${\cal M}_{\rm obs}$ from the ${\cal M}_{219}$ model (red lines) is
compared to marginalized ${\cal D}_{\rm kep}$ (blue histograms and blue lines). The two columns of plots
show the differential (left) and cumulative distribution functions (right). From top to bottom,
the rows of plots show: the (1) radius partitioning ($Q_R$), (2) radius monotonicity ($M_R$), and 
(3) gap complexity (${\cal C}$) (Gilbert \& Fabrycky 2020, He et al. 2020).}
\label{model219c}
\end{figure}

\begin{figure}
\epsscale{0.7}
\plotone{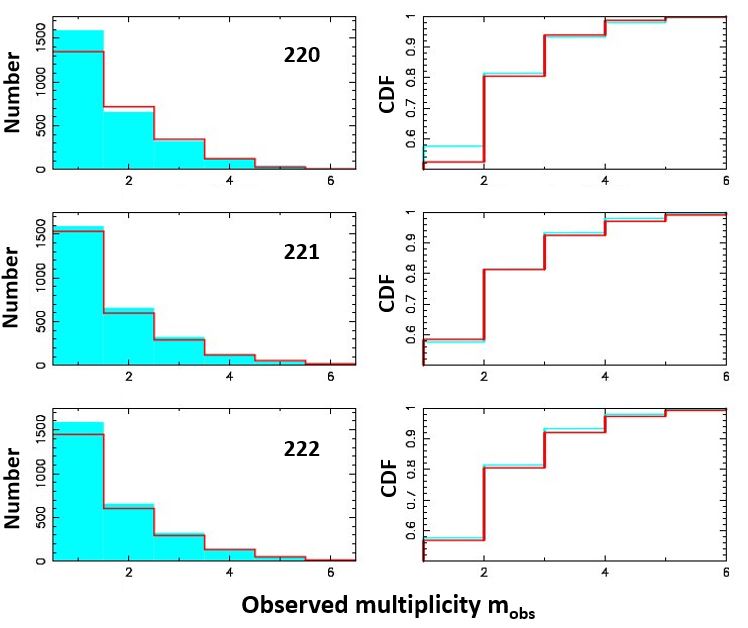}
\caption{Marginalized best-fit observed multiplicity distributions (red lines) from ${\cal M}_{220}$ 
(top panel), ${\cal M}_{221}$ (middle) and ${\cal M}_{222}$ (bottom) are compared to marginalized 
${\cal D}_{\rm kep}$ (blue histogram and blue lines). The two columns of plots show the differential (left) 
and cumulative distribution functions (right).} 
\label{multip}
\end{figure}

Figures \ref{model219a}--\ref{model219c} report the summary statistics from 
${\cal M}_{219}$. We first discuss the parameters that were directly optimized by {\tt MultiNest}.
The multiplicity distribution of Kepler planets is reasonably well reproduced (Fig. \ref{model219a}, 
top). There is a slight tension in ${\cal M}_{219}$, where the optimized fit tends to prefer larger values of 
$\lambda_{\rm pl}$ ($\gtrsim 5$) to fit the observed multiplicity but the total number of detected planets 
by Kepler limits how high $\lambda_{\rm pl}$ can be. This tension increases in a model with $f_{\rm AMD}=0.3$ 
(${\cal M}_{220}$), which struggles to reproduce the observed number of singles (Fig. \ref{multip}). 
The models with the Zipfian multiplicity (${\cal M}_{221}$ and ${\cal M}_{222}$; Fig. \ref{multip}) are less 
susceptible to this problem because they allow for a relatively large fraction of intrinsic singles. 

The optimized models can additionally be influenced by various filters employed in 
Section 2.1. For example, we have $n({\cal D}_{{\rm kep},1})/n({\cal D}_{{\rm kep},2})=1591/657\simeq2.4$ 
(note that here we give values for the {\it planet} multiplicity, as defined in Section 2.2.2), whereas Zhu et al. 
(2018) more aggressively removed suspected FPs among single tranets and found 
$n({\cal D}_{{\rm kep},1})/n({\cal D}_{{\rm kep},2})=432/198\simeq2.2$, 
very close to $n({\cal M}_{{\rm obs},1})/n({\cal M}_{{\rm obs},2}) \simeq 2.1$ in ${\cal M}_{219}$. 
${\cal M}_{219}$ closely matches ${\cal D}_{\rm kep}$ for observed multiplicities $m_{\rm obs} \geq 2$ 
(Fig.~\ref{model219a}). We recall that this not only means that ${\cal M}_{219}$ reproduces the relative
abundance of Kepler planets in systems of different multiplicities -- it matches the {\it actual} 
number of planet detections in ${\cal D}_{{\rm kep},m}$.      

The marginalized orbital period and orbital period ratio distributions in ${\cal M}_{219}$ are statistically 
indistinguishable from Kepler observations (Fig. \ref{model219a}). For example, when all planets are considered,
the Kolmogorov-Smirnov (KS) tests give 33\% and 34\% probabilities for $P_{\rm orb}$ and $P_j/P_{j-1}$, respectively,
that marginalized ${\cal M}_{\rm obs}$ and ${\cal D}_{\rm kep}$ sample the same underlying distribution. 
Interestingly, the orbital period distributions of observed singles and multis in ${\cal D}_{\rm kep}$
differ from each other (third and fourth rows in Fig. \ref{model219a}; the KS test gives only 
a $\sim 10^{-5}$ probability that they are the same), and this difference is quantitatively replicated in 
${\cal M}_{219}$.

Specifically, single tranets tend to have slightly longer orbital periods than multis (e.g., 
$\simeq 32$\% of singles in ${\cal D}_{\rm kep}$ have $P_{\rm orb}>30$ d compared to only $\simeq 23$\% of 
multis). This difference is probably related the Kepler mission baseline (also see Lissauer et al. 2024).    
In ${\cal M}_{219}$, it appears as a consequence of the imposed constraint $P_{\rm orb}<300$ d, where it is 
difficult to fit a long chain of planets within the model domain unless the inner planets have sufficiently 
short orbital periods. Our other base models show the same result for the same reason. 

Additional metrics are plotted in Fig. \ref{model219b}.  We held several model parameters 
-- related to the physical size of planets -- fixed (parameters (7) to (13) in Table 2) to reduce the 
dimensionality of prior space in ${\cal M}_{219}$. The values of these parameters were adopted 
from previous publications and were subsequently adjusted to produce acceptable fits. Indeed, there is 
a very good agreement between the model and data in all metrics shown in Fig. \ref{model219b} (KS test 
probabilities $p>0.05$ in all metrics). 

The last row of in Fig. \ref{model219b} shows the period-normalized transit duration ratio defined as 
$\xi=(T_{{\rm dur},j}/T_{{\rm dur},j-1})(P_j/P_{j-1})^{1/3}$, where $j-1$ and $2 \leq j \leq m_{\rm obs}$ denote the inner 
and outer (detected) planets in a system of observed multiplicity $m_{\rm obs} \geq 2$ (Steffen et al. 2010,
Fabrycky et al. 2014). The parameter $\xi$ is expected to peak near one for planets orbiting the 
same star. The width of the $\xi$ distribution peak increases with increasing mutual inclinations 
of planets. He et al. (2020) pointed out that the maximum AMD model implies lower mutual 
inclinations of systems with higher multiplicities. This produces a narrower $\xi$ peak for higher 
$m_{\rm obs}$ values in ${\cal M}_{219}$, just as observed.
  
Gilbert \& Fabrycky (2020) defined several additional metrics (also see Kipping 2018). The radius partitioning $Q_R$, analogous
to mass partitioning in Gilbert \& Fabrycky (2020), expresses the similarity of planet radii in the same 
system. It is a system-wide characteristic that goes beyond the radius (or depth) ratios of neighbor
planets. The exact definition of $Q_R$ is given in Eqs. (35) and (36) in He et al. (2020): $Q_R=0$ if all 
planets in the same system have the same radius, and $Q_R$ approaches 1 for a system with one dominant 
planet and $m-1$ tiny planets. The radius monotonicity $M_R$, again in close correspondence to the mass
monotonicity in  Gilbert \& Fabrycky (2020), was defined in Eq. (37) of He et al. (2020). This parameter
captures the degree by which the planets are ordered by radius. It is zero for systems that show no 
ordering of planet radius with the orbital period, minus one for perfectly negative monotonic systems,
and plus one for perfectly positive monotonic systems. Additionally, the gap complexity ${\cal C}$ is 
defined in Eqs. (13) and (14) in Gilbert \& Fabrycky (2020). It takes values between 0 and 1, where 
${\cal C}=0$ stands for a system where planets are evenly spaced in log-period and ${\cal C}=1$ 
corresponds to the maximum complexity measure.

Fig. \ref{model219c} shows the $Q_R$, $M_R$ and ${\cal C}$ distributions for ${\cal M}_{219}$. The 
agreement between ${\cal D}_{\rm kep}$ and ${\cal M}_{219}$ is good for $Q_R$ (KS test $p=0.23$).
This shows that the method to generate planetary radii, described in Section 2.2.3, works well. The 
monotonicity parameter $M_R$ in ${\cal M}_{219}$ is skewed toward positive values in much the same way 
the Kepler observations are. As there is no intrinsic radius-period ordering in ${\cal M}_{219}$ (Section 2.2), 
the skewness of $M_R$ arises from our planet detection algorithm: more distant planets have fewer transits, lower 
SNR, and must be larger to be detected. This effect appears to reasonably well explain Kepler observations (KS 
test $p=0.078$), but perhaps not fully as there appears to be a slight excess of planetary systems in 
${\cal M}_{219}$ with $M_R<0$ (Fig. \ref{model219c}, middle plots).
He et al. (2020) also found that the observational bias skews $M_R$ toward positive 
values, but much less strongly than we find here. Conversely, Gilbert \& Fabrycky (2020) showed that 
the observational bias in the model of Mulders et al. (2018) satisfactorily reproduced the monotonicity 
distribution of Kepler planets. A detailed investigation of these subtle differences is left for future work.
 
The gap complexity is the most difficult parameter to match in the population models (Mulders et al. 2018, 
Gilbert \& Fabrycky 2020, He et al. 2020). Similarly to these prior models, ${\cal M}_{219}$ produces a 
relatively broad distribution of ${\cal C}$ that is clearly in tension with the Kepler data (bottom plots 
in Fig.~\ref{model219c}). The gap complexity is influenced by the planet spacing in ${\cal M}_{219}$ 
(depends on the parameter $\sigma_{\rm H}$; we find $\sigma_{\rm H} \simeq 0.36$ in ${\cal M}_{219}$) 
and by non-detections that create artificial gaps between the detected neighbors. 
The gap complexity would decrease for smaller values of $\sigma_{\rm H}$, but that would
also affect the orbit period ratio distribution (Fig.~\ref{model219a}), and the combined constraint 
is difficult to satisfy. The effect of non-detections in ${\cal C}$ is large when AMD is large because 
planets in the same system have relatively large mutual inclinations and may not satisfy the geometric 
condition for transit, creating artificial gaps. Indeed, we obtain slightly better results with 
$f_{\rm AMD}=0.3$ in ${\cal M}_{220}$. 

\begin{figure}
\epsscale{0.8}
\plotone{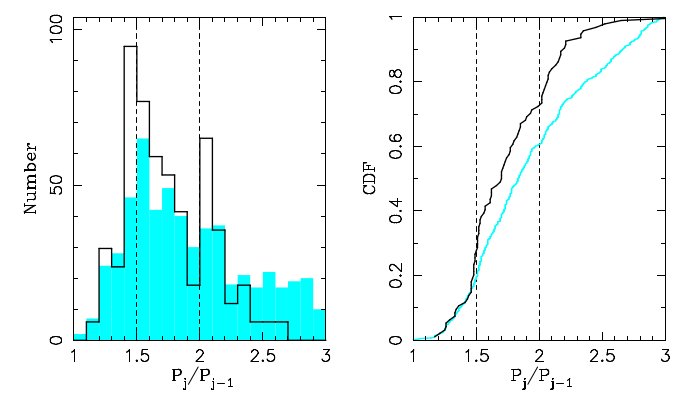}
\caption{The orbital period ratios of all Kepler planets (blue histogram and blue line) are compared to that
of Kepler planets in systems with the gap complexity $C<0.1$ and observed multiplicity $m_{\rm obs} \geq 4$ 
(black lines). Planets in the low gap complexity systems are often found near the 3:2 and 2:1 mean motion 
resonances (vertical dashed lines). These systems contribute to the divergence between the model and observed 
gap complexities in Fig. \ref{model219c}.}
\label{lowc}
\end{figure}

There is a notable excess -- relative to ${\cal M}_{219}$ -- of Kepler systems with very low gap complexities 
in Fig. \ref{model219c}. For example, nearly 60\% of Kepler systems have ${\cal C}<0.1$ compared to only 
about 40\% in ${\cal M}_{219}$. To understand this issue in more detail, we select the Kepler systems 
with ${\cal C}<0.1$ and plot their period ratio distribution ($P_j/P_{j-1}$) in Fig. \ref{lowc}. We find 
that the low-${\cal C}$ systems often have planets with the period ratios $P_j/P_{j-1} \simeq 1.5$ or 2, 
corresponding to the 3:2 and 2:1 MMRs. Interestingly, if a pair of Kepler planets in a system resides near 
the 3:2 MMR, the other planets in the same system show an affinity to the 3:2 MMR as well. A good example 
of this behavior is the KOI-82 (Kepler-102) system with four pairs of planets all having 
$1.33 < P_j/P_{j-1} < 1.7$. We do not obtain these systems often enough in ${\cal M}_{219}$ because there 
are no means built in the base models to produce highly correlated period ratios. The Kepler systems with ${\cal C}<0.1$, 
listed in Table 3, may have retained a memory of their formation conditions. We address this issue in Section 3.2.

${\cal M}_{220}$ with the Poisson multiplicity and $f_{\rm AMD}=0.3$ (Table 4) 
shows results similar to ${\cal M}_{219}$. The model struggles with the multiplicity constraint, producing 17\% fewer 
singles than observed (Fig. \ref{multip}). The split of the orbital period distribution between singles and multis 
is perfect. The gap complexity is slightly improved in ${\cal M}_{220}$ over ${\cal M}_{219}$, because fewer 
artificial gaps are introduced by planet non-detections when $f_{\rm AMD}=0.3$, but a large discrepancy persists. 
The best-fit parameters of ${\cal M}_{220}$ (Table 4) are very similar to those obtained for ${\cal M}_{219}$ 
(Table 2). ${\cal M}_{220}$ prefers slightly larger $P_{\rm break}$ and $f_{\rm H}$ values. 

\begin{figure}
\epsscale{0.8}
\plotone{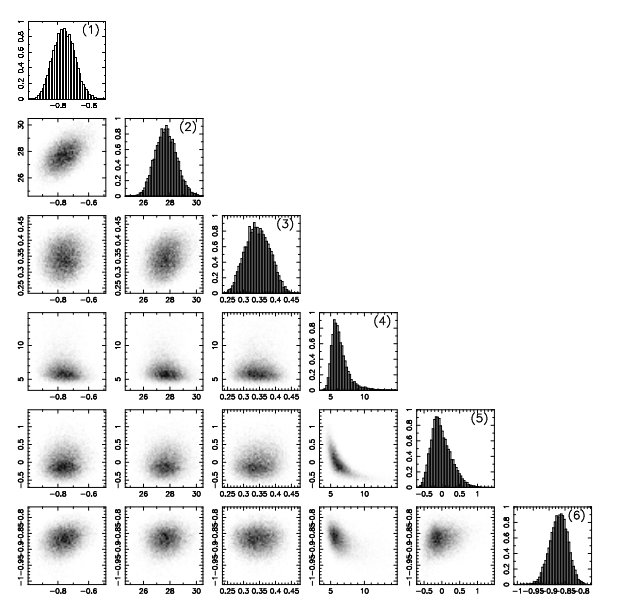}
\caption{A corner plot of posteriors from the ${\cal M}_{221}$ model. See Table 5 for the definition 
of parameters (1) to (6), their prior bounds, and the best-fit values obtained from {\tt MultiNest}.}
\label{corner3}
\end{figure}

${\cal M}_{221}$ with the Zipfian multiplicity and $f_{\rm AMD}=1$ (Fig. \ref{corner3} and Table 5) fits the observed 
multiplicity distribution better than ${\cal M}_{219}$ and ${\cal M}_{220}$ (Fig. \ref{multip}), but is otherwise 
comparable to them. The preferred power-law index for the multiplicity distribution (Eq. \ref{zipfian})
is $\beta_{\rm zipf}=-0.76 \pm 0.07$, indicating a considerably shallower slope than 
$\beta_{\rm zipf} = -1.86 \pm 0.3$ 
found in Sandford et al. (2019). There are more intrinsic singles with the Zipfian 
multiplicity than with the Poisson multiplicity, which helps to more easily 
reproduce the observed multiplicity distribution. ${\cal M}_{222}$ with the Zipfian multiplicity and $f_{\rm AMD}=0.3$ 
(Table 6) gives $\beta_{\rm zipf}=-0.93 \pm 0.08$, shows a bimodal distribution of $P_{\rm break}$, and does not fully 
replicate the different orbital period distributions of singles and multis.  

In summary, all our base models can plausibly reproduce the Kepler data. The observed multiplicity is best 
fit with the Zipfian distribution and $f_{\rm AMD}=1$ (${\cal M}_{221}$), but the other models cannot be ruled 
out (see the Bayes factor differences in Table 1). These results support the finding of Tremaine \& Dong 
(2012) that the Kepler transit photometry alone does not uniquely constrain the multiplicity distribution 
(and other characteristics) of underlying systems. 

It is notable that the relatively simple method 
to generate planetary systems described in Section 2.2 is capable of accurately reproducing many observed 
characteristics (Figs. \ref{model219a}-\ref{multip}). 
This suggests certain homogeneity of the underlying 
population of Kepler planets, and probably reflects common physical mechanisms involved in their formation.      
The gap complexity is the most difficult criterion to satisfy in our base models. There appears to be a 
class of Kepler systems with $m_{\rm obs} \geq 4$, where different planets in the same system 
have highly correlated orbital period ratios $P_j/P_{j-1}$, including a fraction of systems near the 3:2 and 
2:1 MMRs (Fig. \ref{lowc}). Our method to define orbital periods in the base models (Section 2.2.5) does not 
generate these systems often enough. 
     
\subsection{Auxiliary models} 

Here we address the following question: What is the simplest model that would reproduce the measured  
gap complexity while simultaneously preserving good matches in all other metrics? Choosing 
$f_{\rm H}$ from a model with globally fixed $\mu_{\rm H}$ and $\sigma_{\rm H}$ values does not work. If 
$\sigma_{\rm H}$ is large, the gap complexity ends up to be large as well. If $\sigma_{\rm H}$ is small, 
models produce low gap complexities, but the period ratio distribution is wrong (not enough cases 
with $P_j/P_{j-1}<2$). Also, $f_{\rm H}$ cannot be chosen and kept fixed when distributing planets in the 
same system as this would produce extremely low gap complexities (even with $f_{\rm AMD}=1$). The gap 
complexity decreases for lower $f_{\rm AMD}$ values as there are fewer gaps left by non-transiting 
planets. This argument could be used to favor lower $f_{\rm AMD}$ values.

After some testing, we obtained a reasonable fit to the observed gap complexity distribution when we 
assumed that the systems with the intrinsic multiplicity 
$m_{\rm int} < m^*$ have uncorrelated orbital period ratios (i.e., when the 
$f_{\rm H}$ factors for different planets in the same system are randomly chosen from a Gaussian 
distribution; Section 2.2.5), and $m_{\rm int} > m^*$ have ideally correlated period ratios (i.e., the same 
$f_{\rm H}$ factors for all planets in the same system). The best-fit transition multiplicity $m^*$
was determined by the following method. First, we modified the base models to include the transition
to strictly correlated period ratios at $m^*$. Second, we set $m^*=2.5$, 3.5, 4.5 and 5.5 in different 
auxiliary models to explore different options. Third, following the procedure outlined in Section 2.4,
we evaluated the likelihood term in $(n({\cal M}_{{\rm obs},m}), P_{\rm orb})$, 
$(n({\cal M}_{{\rm obs},m}),P_j/P_{j-1})$, and $(n({\cal M}_{{\rm obs},m}),{\cal C})$. Here, ${\cal C}$ is 
explicitly included in the third likelihood array such that our auxiliary models can address the gap complexity problem. 
The auxiliary models have seven priors: the same six priors as the base models, $\lambda_{\rm pl}$ (Poisson 
multiplicity),\footnote{We also tested several auxiliary models with the Zipfian multiplicity and found
that these models do not constrain $m^*$ or $f_{\rm AMD}$ well. Specifically, they are indiscriminate as 
for the value of the transition multiplicity $m^*$, and favor $f_{\rm AMD} \sim 1$. We do not find these 
results useful. Moreover, all these models are disfavored by a large Bayes factor 
($\Delta \ln {\cal Z} < -30$) with respect to ${\cal M}_{228}$.} $f_{\rm H}$, 
$\sigma_{\rm H}$, $P_{\rm break}$, $\alpha_P$, $\beta_P$, plus $f_{\rm AMD}$. We include $f_{\rm AMD}$ because 
we find that this parameter can be meaningfully constrained when ${\cal C}$ is used in the likelihood 
term.\footnote{This cannot be done in our base models, where ${\cal C}$ is not 
used in the likelihood term. When we include $f_{\rm AMD}$ as a prior parameter in the base models, 
the modified fit requires that $f_{\rm AMD} \sim 1$, as needed to strictly reproduce the observed 
multiplicity distribution. We do not consider this result particularly meaningful for reasons discussed 
in Section 3.1.}  

\begin{figure}
\epsscale{0.8}
\plotone{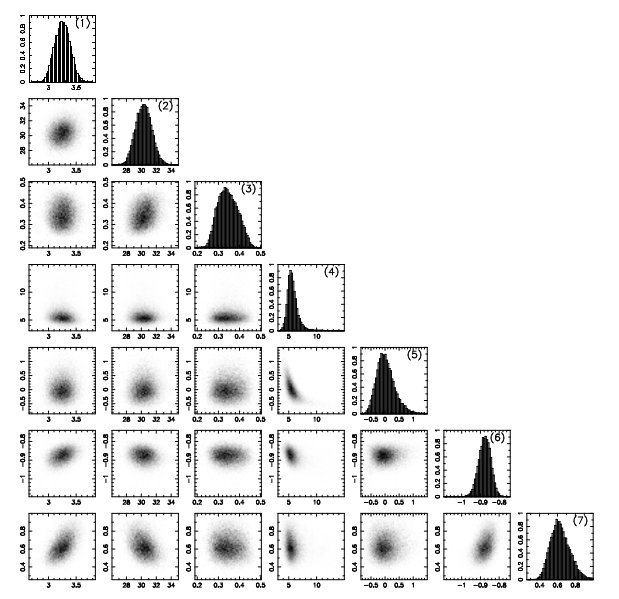}
\caption{A corner plot of posteriors from the ${\cal M}_{228}$ model. See Table 7 for the definition 
of parameters (1) to (7), their prior bounds, and the best-fit values obtained from {\tt MultiNest}.}
\label{corner2}
\end{figure}

Our auxiliary models with $m^*=2.5$ and 5.5 are strongly disfavored compared to $m^*=4.5$ ($\Delta \ln {\cal Z} < -17$;
Table 1). The model with $m^*=3.5$ is also disfavored with $\Delta \ln {\cal Z} = -7.1$. We therefore focus on 
${\cal M}_{228}$ with $m^*=4.5$, where the transition to strictly correlated orbital radii happens between
intrinsic multiplicities 4 and 5. Figure \ref{corner2} shows a corner plot from ${\cal M}_{228}$ and Table
7 reports the best fit parameters and their uncertainties. The posterior distribution in Fig. \ref{corner2}
is well behaved. We find $f_{\rm AMD}=0.62^{+0.12}_{-0.10}$, an intermediate value between $f_{\rm AMD}=0.3$ and 
1 explored in the base models (Section 3.1). This is significant because $f_{\rm AMD}$ has not been constrained in 
the base models, that is without including the gap complexity in likelihood. For comparison, He et al. (2020) 
reported that $f_{\rm AMD}$ peaks at $\sim 0.5$ in the planet formation simulations of Carrera et al. (2018),
with a tail to larger values. 

Figure \ref{model228} shows the summary statistics from ${\cal M}_{228}$. There is a very good correspondence 
between the model and observations for all parameters (the KS test probabilities $p>0.1$ in all metrics).
The agreement for other parameters, not shown here for brevity, is as good as in models 
${\cal M}_{219}$ (Fig. \ref{model219b} and \ref{model219c}).\footnote{Note that the gap complexity is included
in the likelihood term in ${\cal M}_{228}$, but not in the base models. We are therefore unable to 
statistically compare the base and auxiliary models based on the Bayes factor differences. 
${\cal M}_{228}$ is preferred over the base models because it better fits the gap complexity parameter.}  
The most significant improvement over the base models is in the gap complexity distribution (compare the bottom panel in 
Fig. \ref{model219c} from ${\cal M}_{219}$ with Fig. \ref{complex} from ${\cal M}_{228}$). The new model not 
only adequately matches the observed ${\cal C}$ distribution for all Kepler planets with $m_{\rm obs} \geq 3$
-- it also separately reproduces the distinct ${\cal C}$ distributions of systems with $m_{\rm obs} = 3$
and $m_{\rm obs} \geq 4$.

The gap complexity distributions for $m_{\rm obs}=3$ and $m_{\rm obs} \geq 4$ in ${\cal D}_{\rm kep}$ are significantly 
different (Fig. \ref{complex}) with larger multiplicities showing greater fraction of low complexity systems. 
In ${\cal M}_{228}$, which correctly mimics this difference, this is a consequence of the transition to 
highly correlated orbital periods at $m^*=4.5$. A large fraction of systems with $m_{\rm obs} = 3$ are derived 
from systems with intrinsic multiplicities $m_{\rm int} = 3$ and 4, which do not have strictly correlated 
orbital periods in the model, whereas the ones with $m_{\rm obs} \geq 4$ are predominantly derived from systems 
with highly correlated orbital periods. If this interpretation is correct, the observed high multiplicity systems 
($m_{\rm obs} \geq 4$) could have retained a memory of their formation conditions, which presumably produced 
correlated orbital radii via accretion and planet migration in a protoplanetary gas disk, and orbital 
resonances between planets. The lower multiplicity systems ($m_{\rm obs} \leq 3$) presumably evolved 
from these initial configurations by dynamical instabilities (e.g., Izidoro et al. 2017, 2021; Goldberg 
\& Batygin 2022).

The radius monotonicity $M_R$ in ${\cal M}_{228}$ shows a dependence on the radius of the innermost planet,
$R_1$, and reproduces the Kepler data well (Fig. \ref{monot}). For $R_1<2$ $R_\oplus$,
there is a strong preference toward positive monotonicities with $\simeq 75$\% of systems having $M_R>0$ 
(also $\simeq 80$\% of Kepler systems have $M_R>0$ for $R_1<1.5$ $R_\oplus$ but the statistics for 
$R_1<1.5$ $R_\oplus$ are poor). This trend nearly vanishes for $R_1>2$ $R_\oplus$, where only 
$\simeq 55$\% of systems have $M_R>0$ (there is no preference for $M_R>0$ if $R_1>2.5$ $R_\oplus$ but the 
statistics for $R_1>2.5$ $R_\oplus$ are poor). We believe that these trends arise from the observational 
bias as the systems with small inner planets need to host relatively large outer planets such that 
these outer planets can be detected by Kepler, $m_{\rm obs} \geq 3$, and $M_R$ can be defined. {In any case, 
given the currently small statistics for $R_1>2$ $R_\oplus$ (only 17 counts; bottom panels in Fig. 
\ref{monot}), the trends tentatively identified here will need to be tested with more future data.}


\subsection{Intrinsic distributions}

Figures \ref{intrin1}-\ref{intrin3} show the {\it intrinsic} distributions obtained in ${\cal M}_{228}$ 
(Poisson, $m^*=4.5$) and ${\cal M}_{235}$ (Zipfian, $m^*=4.5$) -- our two best models in each category 
with seven priors and the gap-complexity constraint (note that the Zipfian model ${\cal M}_{235}$ 
is disfavored relative to ${\cal M}_{228}$ ). The intrinsic planet multiplicity in ${\cal M}_{228}$ favors systems with
$m_{\rm int}=2$-4, whereas in ${\cal M}_{235}$ the distribution is relatively flat for 
$m_{\rm int}\leq3$ and decreases less strongly for higher multiplicities (Fig. \ref{intrin1} top).
The two models give very similar results for the overall period distribution, the period
distribution of multis, and the orbital period ratio, respectively. As for the orbital period of 
singles, however, ${\cal M}_{235}$ has a relatively low proportion of single planets with 
$P_{\rm orb}>100$ d. The singles with $P_{\rm orb}>100$ d are more often represented in ${\cal M}_{228}$, 
because of the Kepler baseline cutoff (planets with $P_{\rm orb}>300$ d are often there with the 
Poisson multiplicity distribution in ${\cal M}_{228}$, but are not included in the count).

\begin{figure}
\epsscale{0.8}
\plotone{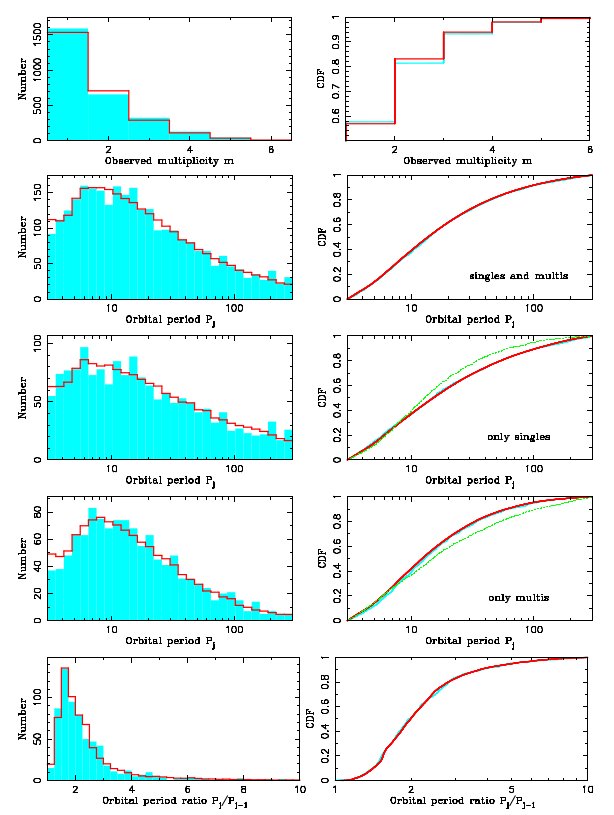}
\caption{Marginalized best-fit ${\cal M}_{\rm obs}$ from the ${\cal M}_{228}$ model (red lines) is
compared to marginalized ${\cal D}_{\rm kep}$ (blue histograms and blue lines). The two columns of plots
show the differential (left) and cumulative distribution functions (right). From top to bottom,
the rows of plots show the: (1) observed multiplicity ($m_{\rm obs}$), (2) orbital periods of all 
planets ($P_{\rm orb}$), (3) orbital periods of singles (planets in systems of observed multiplicity $m_{\rm obs}=1$), 
(4) orbital periods of multis (observed multiplicity $m_{\rm obs} \geq 2$), and (5) orbital period ratios 
of neighbor planets ($P_j/P_{j-1}$ for $2 \leq j \leq m_{\rm obs}$). The green dashed lines in rows (3) and (4) are plotted for reference: 
the period distribution of multis in (3) and singles in (4).}
\label{model228}
\end{figure}

The effect of observational bias in Kepler's transit detections becomes obvious when the biased distributions
in Fig. \ref{model228} are compared with the intrinsic distributions in Fig. \ref{intrin1}
(red lines for ${\cal M}_{228}$). As expected, the Kepler observations strongly favor the detection
of planets with low orbital periods (e.g., $\simeq 70$\% of detected planets with $P_{\rm orb}<30$ d
to be compared to only $\simeq 30$\% of planets with $P_{\rm orb}<30$ d in the intrinsic distribution). 
The observed period ratio distribution (Fig. \ref{model228} bottom) is broader than the intrinsic 
distribution (Fig. \ref{intrin1} bottom), because non-detections produce gaps between the detected 
planets and larger $P_j/P_{j-1}$ ratios in the detected population. Similarly, observational biases
broaden the planetary radius and transit depth distributions, but do not alter the radius ratio 
distribution too much {(compare Fig. \ref{intrin2} to Fig. \ref{model219b})}.

The intrinsic radius partitioning distributions in ${\cal M}_{228}$ and ${\cal M}_{235}$ are very similar
to each other, but ${\cal M}_{228}$ has more systems with $m_{\rm int} \geq 3$ than ${\cal M}_{235}$; ${\cal M}_{228}$ 
therefore plots higher in the differential plots than ${\cal M}_{235}$ {(Fig. \ref{intrin3})}.
The same applies to the radius monotonicity. Notably, the intrinsic radius monotonicity is skewed toward
positive values in Fig. \ref{intrin3} (middle panels). The positive monotonicity is introduced in 
the model for planet chains with relatively small inner planets. When the outer planets are inserted  
in these systems, the positive monotonicity arises when planets with $R < 0.5$ $R_\oplus$ are rejected
(see Section 2.2.3). We verified that the same trend exists in the Kepler dataset ${\cal D}_{\rm kep}$, 
where systems with less massive planets also show stronger positive monotonicities (Fig. \ref{monot}). 

The Zipfian model ${\cal M}_{235}$ with $m^*=4.5$ shows a stronger preference for very low (intrinsic) 
gap complexities than the Poisson model with $m^*=4.5$. This effect arises because the Zipfian model
has more planetary systems with $m_{\rm int}>4.5$ (Fig. \ref{intrin1} top) than the Poisson model, and 
is therefore more affected by the transition to the highly correlated systems with $m_{\rm int}>m^*$.  
Compared to both intrinsic gap complexity distributions shown in Fig. \ref{intrin3}, the biased 
gap complexity distribution in ${\cal M}_{228}$ is wider due to gaps produced by non-transiting 
planets (Fig. \ref{complex}).
     
\section{Planet occurrence}

Here we use ExoMOD to make various predictions about the planet occurrence. Synthesizing the 
results of all models investigated here, we find that a FGK dwarf in the Kepler field should host 
$\simeq 2.4 \pm 0.2$ planets on average with $0.5 < R_{\rm pl}/R_\oplus < 7$ and $3<P_{\rm orb}<300$ 
days. The Poisson multiplicity models ${\cal M}_{219}$ and  ${\cal M}_{220}$ suggest $\simeq 2.5$ 
planets on average, slightly higher than the average of $\simeq 2.3$ planets in the Zipfian multiplicity 
models ${\cal M}_{221}$ and ${\cal M}_{222}$. The auxiliary model ${\cal M}_{228}$ gives $\simeq 2.4$ 
planets on average. In the Poisson models, we have $\lambda_{\rm pl}=3$-3.5, but about 25\% of planets 
generated with the method described in Section 2.2.5 have $P_{\rm orb}>300$ d and are not included here. 

With 107,983 FGK dwarfs in our filtered catalog, this represents 240,000-280,000 planets in total. Of these, 
$\simeq 2.3$\% are transiting and $\simeq 1.0$\% are transiting {\it and} detected by Kepler (i.e., recorded 
in ${\cal M}_{\rm obs}$), which favorably compares with 2761 planets in our input catalog (${\cal D}_{\rm kep}$, 
Section 2.1). Nearly all Kepler stars (97\%) have at least one planet in the model domain in ${\cal M}_{219}$, 
${\cal M}_{220}$ and ${\cal M}_{228}$. In ${\cal M}_{221}$ and ${\cal M}_{222}$, this fraction is 100\% because 
we only consider $m_{\rm int} \geq 1$ in this multiplicity model (Section 2.2.2).

The majority of detected single tranets have undetected companions in the model domain in all our 
models (i.e., they reside in systems with the intrinsic multiplicity $m_{\rm int}\geq2$): 84\% in ${\cal M}_{219}$, 
81\% in ${\cal M}_{220}$, 71\% in ${\cal M}_{221}$, 62\% in ${\cal M}_{222}$, and 85\% in ${\cal M}_{228}$. 
The fraction decreases in the Zipfian multiplicity models because there are more single planets generated in 
these models (especially
if the power index is steeper like in ${\cal M}_{222}$). For example, only $\simeq11$\% and $\simeq 15$\% of 
all generated planets were intrinsic singles in ${\cal M}_{221}$ and ${\cal M}_{222}$, respectively.
The average number of non-detected companions in systems with detected single tranets ranges from 
$\simeq 2.0$ in ${\cal M}_{219}$ down to $\simeq 1.5$ in ${\cal M}_{222}$.

When one or more planets are detected by Kepler around a specific host star, Kepler observations typically 
miss at least one planet with $P<300$ d. The number of missed planets depends on the observed multiplicity.
For example, for $m_{\rm obs}=1$, 2, 3, 4 and 5 in ${\cal M}_{228}$, the mean number of missed planets is 
1.85, 1.68, 1.42, 1.20, 0.72, respectively. The smaller number of missed planets in systems with higher 
multiplicity ${\cal M}_{228}$ arises as a consequence of the AMD model. The great majority of missed planets have longer orbital
periods than the detected planets. If we instead ask how many planets are not detected with orbital radii 
smaller than the outermost detected planet, we find on average that only $0.39$ (inner or intermediate) 
planets are not detected in systems of the observed multiplicity $m_{\rm obs}=5$, and $\simeq 0.51$-0.58 
in systems with $m_{\rm obs}\leq 4$. As a rule of thumb, there is therefore a $\sim50$\% probability that 
an observed Kepler system is missing inner or intermediate-period planets.    
 
A simple definition of the habitable zone is adopted here: $q=a_{\rm pl}(1-e_{\rm pl})>r_{\rm in}$ and 
$Q=a_{\rm pl}(1+e_{\rm pl})<r_{\rm out}$ with $r_{\rm in}=\sqrt{L_*/1.1}$ and $r_{\rm out}=\sqrt{L_*/0.53}$,
where $L_*$ is the stellar luminosity. We find that a fraction $\eta=0.28$ of stars have a planet with
$0.5 < R_{\rm pl}/R_\oplus < 7$ in the habitable zone in ${\cal M}_{228}$. If we further restrict 
the selection to planet radii $0.5 < R_{\rm pl}/R_\oplus < 2$, we find the eta-Earth $\eta_\oplus = 0.17$, 
indicating that about 17\% of Kepler stars host a habitable planet in ${\cal M}_{228}$. 
Our models with the Poisson multiplicity show similar eta-Earth factors: $\eta_\oplus = 0.18$ in 
${\cal M}_{219}$, and $\eta_\oplus = 0.15$ in ${\cal M}_{220}$. We conclude that 15-18\% of Kepler stars 
should host a planet with $0.5 < R_{\rm pl}/R_\oplus < 2$ in the habitable zone (with the definition 
above). The Zipfian models indicate more varied values: $\eta_\oplus = 0.28$ in ${\cal M}_{221}$, and 
$\eta_\oplus = 0.16$ in ${\cal M}_{222}$.  

{For comparison, Mulders et al. (2018) defined the habitable-zone planets as $0.9 < P/P_{\oplus} <2.2$
and $0.7 < R/R_{\oplus} <1.5$, where $P_\oplus$ is the orbital period of the Earth, and found 
$\eta_\oplus = 36 \pm 14$ \%. We obtain $\eta_\oplus = 12$\% in ${\cal M}_{228}$ by adopting the same criteria, 
about a factor three below Mulders et al. (2018), but only 5\% lower than what we obtained in ${\cal M}_{228}$ 
with our definition. This shows that the main difference in $\eta_\oplus$ is the population model itself, rather 
than the definition of the habitable zone. Hsu et al. (2019) estimated $\eta_\oplus = 16^{+11}_{-6}$\% for $0.65 < P/P_{\oplus} 
< 1.37$ and $0.75 < R/R_{\oplus} <1.5$, which is more in line with our findings (we compute $\eta_\oplus=13$\% 
for the same definition). Eta-Earth could be higher if a more optimistic definition of the habitable zone
is adopted and/or if the detection efficiency is lower in higher multiplicity systems (e.g., Zink et al. 
2019).}

We did not implement any orbital radius cutoff, other than $q>r_{\rm in}$ and $Q<r_{\rm out}$, when reporting 
the eta-Earth factors above. Since our models were calibrated on Kepler planets with $P_{\rm orb} < 300$ d,
to calculate $\eta_\oplus$, we assumed that there is no change in the orbital architecture of planets at 
$P_{\rm orb} \sim 300$ d. Millholland et al. (2022) found that the Kepler systems tend to be confined 
to $P_{\rm orb}<100$-300 d, and that there is paucity of planets beyond that edge. To crudely account for 
the edge effect on $\eta_\oplus$, we repeated the eta-Earth calculation schematically assuming that there 
are no planets with $P_{\rm orb}>300$ d and found that the edge effect could be profound. For example, in ${\cal M}_{228}$,
where we previously had $\eta_\oplus = 0.17$, the new calculation gives $\eta_\oplus = 0.05$. A better 
observational characterization of Earth-size planets at large orbital radii is obviously needed.

\section{Conclusions}

The main results of this work are summarized as follows.

\begin{figure}
\epsscale{0.7}
\plotone{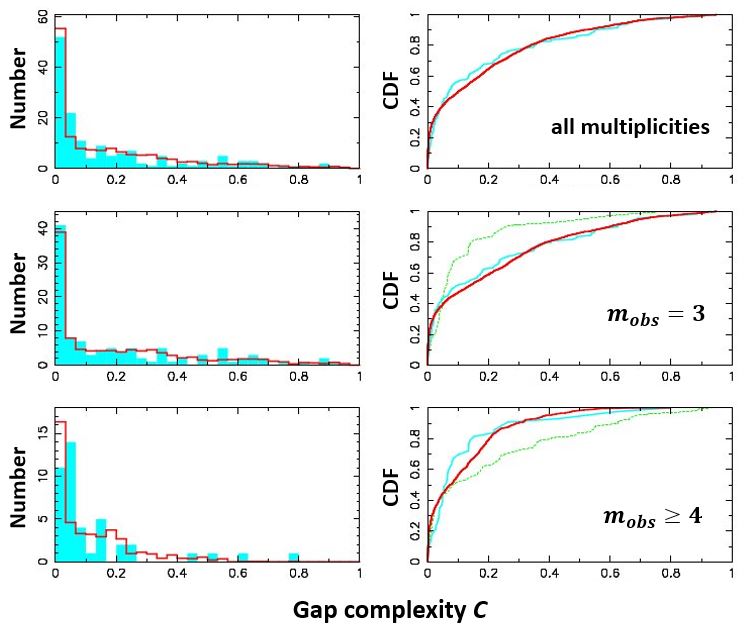}
\caption{Marginalized best-fit ${\cal M}_{\rm obs}$ from the ${\cal M}_{228}$ model ($m^*=4.5$, red lines) 
is compared to marginalized ${\cal D}_{\rm kep}$ (blue histograms and blue lines). The two columns of plots
show the differential (left) and cumulative distribution functions (right). From top to bottom,
the rows of plots show: the (1) gap complexity ${\cal C}$ for all Kepler systems (Gilbert \& 
Fabrycky 2020, He et al. 2020), (2) ${\cal C}$ for systems with $m_{\rm obs}=3$, and (3) ${\cal C}$ for 
systems with $m_{\rm obs} \geq 4$. The green lines in rows (2) and (3) are plotted for 
reference: the ${\cal C}$ distribution for $m_{\rm obs} \geq 4$ in (2) and $m_{\rm obs}=3$ in (3).} 
\label{complex}
\end{figure}

\begin{figure}
\epsscale{0.7}
\plotone{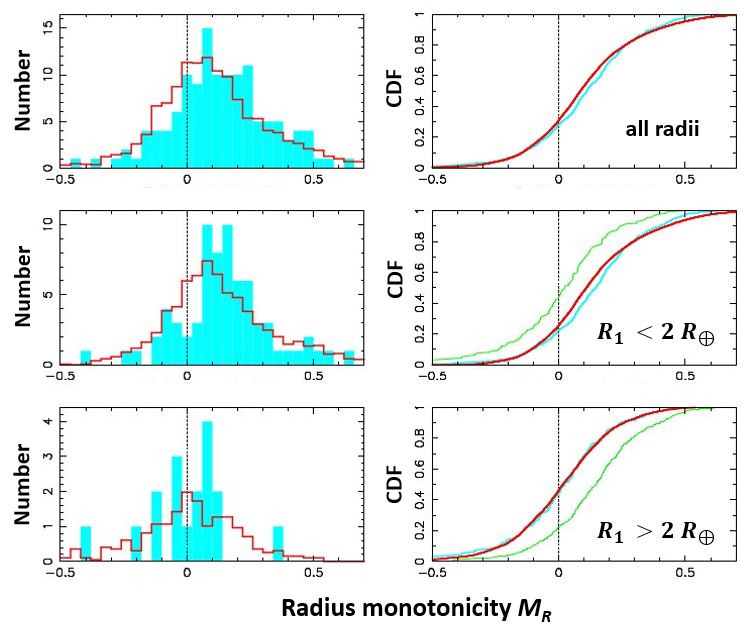}
\caption{Marginalized best-fit ${\cal M}_{\rm obs}$ from the ${\cal M}_{228}$ model ($m^*=4.5$, red lines) 
is compared to marginalized ${\cal D}_{\rm kep}$ (blue histograms and blue lines). The two columns of plots
show the differential (left) and cumulative distribution functions (right). From top to bottom,
the rows of plots show: the (1) radius monotonicity $M_R$ for all Kepler systems (Gilbert \& 
Fabrycky 2020, He et al. 2020), (2) $M_R$ for systems with $R_1<2$ $R_\oplus$, and (3) $M_R$ for 
systems with $R_1>2$ $R_\oplus$. The green lines in rows (2) and (3) are plotted for 
reference: the $M_R$ distribution for $R_1 > 2$ $R_\oplus$ in (2) and $R_1 > 2$ $R_\oplus$ in (3).} 
\label{monot}
\end{figure}

\begin{figure}
\epsscale{0.8}
\plotone{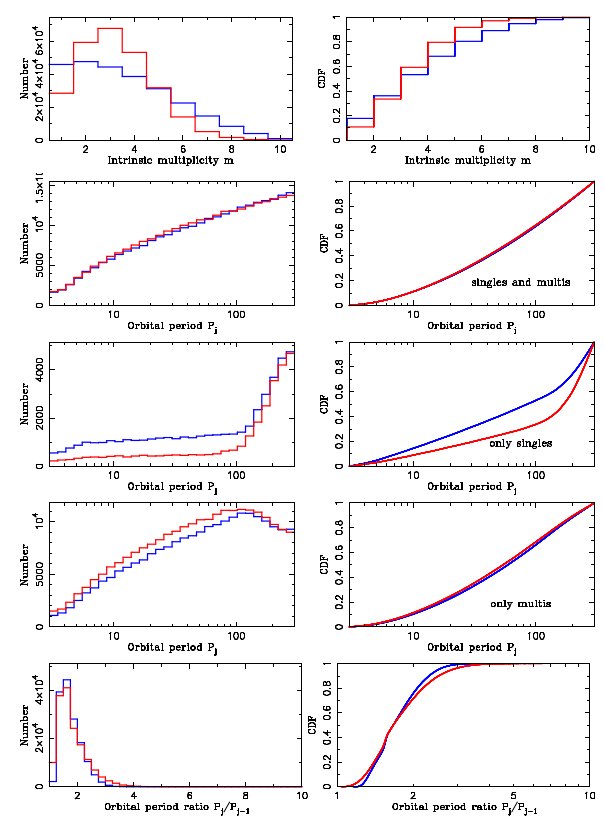}
\caption{Marginalized best-fit ${\cal M}_{\rm int}$ from the ${\cal M}_{228}$ (Poisson, red lines) and 
${\cal M}_{235}$ (Zipfian, blue lines) models. The two columns of plots
show the differential (left) and cumulative distribution functions (right).
From top to bottom, the rows of plots show the: (1) intrinsic multiplicity ($m_{\rm int}$), (2) orbital periods 
of all planets ($P_{\rm orb}$), (3) orbital periods of singles (planets in systems of intrinsic multiplicity 
$m_{\rm int}=1$), (4) orbital periods of multis (intrinsic multiplicity $m_{\rm int} \geq 2$), and (5) orbital 
period ratios of neighbor planets ($P_j/P_{j-1}$ for $2 \leq j \leq m_{\rm int}$).  Note that the top plots 
show the planet multiplicity, as defined in Section 2.2.2, and not the system multiplicity as in Zhu et al. 
(2018) or He et al. (2020).}
\label{intrin1}
\end{figure}

\begin{figure}
\epsscale{0.8}
\plotone{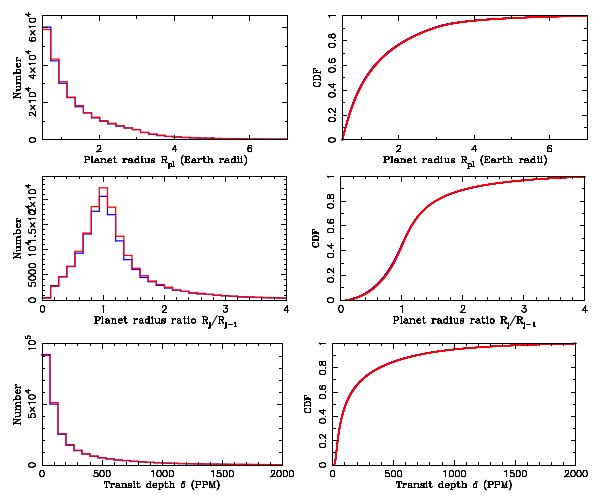}
\caption{Marginalized best-fit ${\cal M}_{\rm int}$ from the ${\cal M}_{228}$ (Poisson, red lines) and 
${\cal M}_{235}$ (Zipfian, blue lines) models. The two columns of plots
show the differential (left) and cumulative distribution functions (right).
From top to bottom, the rows of plots show: the (1) planet radius ($R_{\rm pl}$), (2) radius ratio of 
neighbor planets ($R_j/R_{j-1}$ for $2 \leq j \leq m_{\rm int}$), (3) transit depth ($\delta$),}
\label{intrin2}
\end{figure}

\begin{figure}
\epsscale{0.8}
\plotone{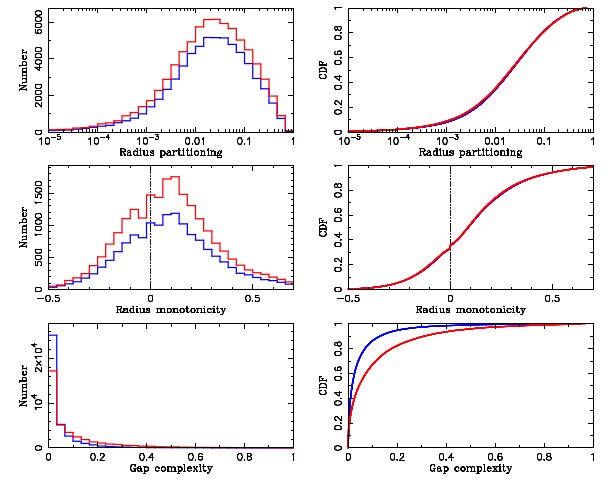}
\caption{Marginalized best-fit ${\cal M}_{\rm int}$ from the ${\cal M}_{228}$ (Poisson, red lines) and 
${\cal M}_{235}$ (Zipfian, blue lines) models. The two columns of plots
show the differential (left) and cumulative distribution functions (right).
From top to bottom,
the rows of plots show: the (1) radius partitioning ($Q_R$), (2) radius monotonicity ($M_R$), and 
(3) gap complexity (${\cal C}$) (Gilbert \& Fabrycky 2020, He et al. 2020).}
\label{intrin3}
\end{figure}

\begin{enumerate}
\item The population model developed here, ExoMOD, is capable of reproducing many characteristics of close-in 
Kepler planets. This indicates certain homogeneity of the underlying population of Kepler planets, and probably 
reflects common physical process involved in their formation (disk-driven migration, capture in
orbital resonances, dynamical instabilities, etc.). 
\item The orbital period distributions of Kepler singles and multis are statistically different from 
each other (this applies to both the observed, Section 3, and intrinsic distributions, Section 4). 
The difference is quantitatively reproduced in ExoMOD as a consequence of the imposed constraint, $P_{\rm orb}<300$ d, 
as motivated by the Kepler mission baseline. 
\item Our base models do not match the gap complexity distribution of 
Kepler planets (${\cal C}$, Gilbert \& Fabrycky 2020). 
A relatively large fraction of Kepler's low-${\cal C}$ systems have highly correlated orbital 
period ratios suggesting a regular radial spacing of planets. To account for this characteristics, we 
modified the base models to include (ideally) correlated period ratios for the intrinsic multiplicities 
$m_{\rm int} \geq 5$. The modified models match the ${\cal C}$ distribution of Kepler systems with 
$m_{\rm obs} \geq 3$, and separately for $m_{\rm obs}=3$ and $m_{\rm obs} \geq 4$ (Fig. \ref{complex}).
Higher-multiplicity planetary systems with low gap complexities (Table 3) probably retained a memory of their 
formation conditions.
\item The Kepler transit photometry alone does not uniquely constrain the multiplicity and/or AMD distribution 
of the underlying systems in our base models. The situation improves when the gap complexity is included 
as a constraint in the likelihood term. Our best auxiliary model, ${\cal M}_{228}$, favors the Poisson
multiplicity with $\lambda_{\rm pl}=3.25\pm0.14$ (the average number of planet without cuts) and 
$f_{\rm AMD} \simeq 0.62 \pm 0.12$ (the fraction of maximum AMD distributed among planets). 
\item We find that systems with more positive radius monotonicities ($M_R>0$) typically harbor smaller planets, 
  as expected if the radius monotonicity is influenced by non-detection of small planets. For $R_1<2$ $R_\oplus$,
  there is a strong trend toward positive monotonicities with $\simeq 75$\% of systems having $M_R>0$.
  For $R_1>2$ $R_\oplus$, only $\simeq 55$\% of systems have $M_R>0$. 
\item A FGK dwarf in the Kepler field should host $\simeq 2.4 \pm 0.2$ planets on average with $0.5 < R_{\rm pl}/R_\oplus < 7$ 
and $3<P_{\rm orb}<300$ days. The majority (60-85\%) of detected single tranets have undetected companions. 
Nearly all Kepler stars ($>95$\%) should have at least one planet with the orbital period $3<P_{\rm orb}<300$ days
and physical radius $0.5 < R_{\rm pl}/R_\oplus < 7$.  For a detected planetary system, there is a $\sim50$\% 
probability that Kepler transit observations missed at least one inner or intermediate-period planet.  
\item 
For a simple definition of the habitable zone adopted here (Section 4), and assuming that there is no change in 
the orbital architecture of planets from $P_{\rm orb} <300$ d to $P_{\rm orb} >300$ d, we estimate that 15-18\% 
of Kepler stars should host a planet with $0.5 < R_{\rm pl}/R_\oplus < 2$ in the habitable zone (eta-Earth 
$\eta_\oplus=0.15$-0.18). When we crudely account for the possible edge effect (Millholland et al. 2022) by
removing all planets with $P_{\rm orb} >300$ d, ExoMOD gives $\eta_\oplus \simeq 0.05$. A better characterization 
of Earth-size planets at large orbital radii is needed to anchor eta-Earth. 
\end{enumerate} 

ExoMOD can be used to produce random samples of intrinsic and Kepler-detected planetary systems. Ten samples from 
the ${\cal M}_{228}$ model, generated with different random seeds, can be found at {\tt www.boulder.swri.edu/\~{}davidn/ExoMOD/}. 
{These samples can be useful for a number of scientific projects. For example, they can be used to predict properties 
of missing planets (Dietrich \& Apai 2020) or help to constrain dynamical models of Kepler planet formation 
(e.g., Izidoro et al. 2021).} Additional model information is available upon request.  

\clearpage

\acknowledgements

\begin{center}
{\bf Acknowledgments} 
\end{center} 
\vspace*{-3.mm}
The simulations were performed on the NASA Athena Supercomputer. We thank the NASA NAS computing division for 
continued support. D.N.'s work was funded by the NASA XRP program. D.A.Y's work is supported by a Juan Carlos 
Torres Postdoctoral Fellowship at the Massachusetts Institute of Technology. {We thank the anonymous reviewer 
for their insightful comments on the submitted manuscript.}


\begin{table}
\centering
{
\begin{tabular}{lcccc}
\hline \hline
Model           & Multiplicity      &  $f_{\rm AMD}$  & $m^*$    & $\Delta \ln {\cal Z}$   \\                      
\hline                    
${\cal M}_{219}$ & Poisson           &  1.0       & --        & -0.5            \\ 
${\cal M}_{220}$ & Poisson           &  0.3       & --        & -3.6            \\ 
${\cal M}_{221}$ & Zipfian           &  1.0       & --        &  0            \\ 
${\cal M}_{222}$ & Zipfian           &  0.3       & --        & -2.9            \\
\hline
${\cal M}_{230}$ & Poisson           &  --        & 2.5       & -17.4         \\ 
${\cal M}_{229}$ & Poisson           &  --        & 3.5       & -7.1          \\ 
${\cal M}_{228}$ & Poisson           &  --        & 4.5       &  0            \\ 
${\cal M}_{231}$ & Poisson           &  --        & 5.5       & -17.1          \\ 
\hline \hline
\end{tabular}
}
\caption{A summary of selected models. The columns are: the (1) model designation, (2) adopted intrinsic multiplicity 
distribution, (3) fraction $f_{\rm AMD}$ of the critical AMD distributed among planets, (4) 
multiplicity transition $m^*$ to highly correlated orbital radii, and (5) Bayes factor 
difference relative to the best model in each category. The four models in the top part of 
the table are our base models with six priors and two likelihood arrays (Section 3.1).
The four models in the bottom part of the table are our auxiliary models with seven priors 
and three likelihood arrays (Section 3.2).
}
\end{table}

\begin{table}
\centering
{
\begin{tabular}{lcccc}
\hline \hline
    & Parameter        & Units      &  Value             &  Prior bounds          \\                      
\hline                    
(1) & $\lambda_{\rm pl}$  & --        & $3.42^{+0.12}_{-0.12}$       & $(1,7)$     \\ 
(2) & $f_{\rm H}$        & --         & $26.69^{+0.77}_{-0.76}$      & $(15,35)$  \\ 
(3) & $\sigma_{\rm H}$   & --         & $0.3646^{+0.042}_{-0.040}$    & $(0,1.5)$  \\           
(4) & $P_{\rm break}$     & days       & $5.16^{+1.11}_{-0.76}$       & $(3,15)$   \\           
(5) & $\alpha_P$        & --         & $-0.077^{+0.394}_{-0.305}$   & $(-1,1.5)$  \\  
(6) & $\beta_P$         & --         & $-0.796^{+0.023}_{-0.027}$   & $(-1.5,0)$   \\
\hline
(7) & $f_{\rm AMD}$       & --         & 1.0                     &  --        \\
(8) & $R_{\rm break}$     & $R_\oplus $  & $3.0$                    &  --        \\  
(9) & $\alpha_R$        & --         & $-1.5$                  & --         \\  
(10) & $\beta_R$         & --         & $-8.0$                  & --         \\
(11) & $\sigma_R$       & --         & $0.22$                  & --         \\
(12) & $\nu_R$          & --         & $0.5$                   & --         \\  
(13) & $\sigma_e$       & --          & $0.25$                  & --         \\ 
\hline \hline
\end{tabular}
}
\caption{Model ${\cal M}_{219}$ with the Poisson multiplicity distribution and $f_{\rm AMD}=1$. 
The prior bounds of parameters (1) to (6) are given in the last column. Parameters (7) to (13) 
were fixed in ${\cal M}_{219}$ to the values given in the fourth column. }
\end{table}

\begin{table}
\centering
{
\begin{tabular}{lccc}
\hline \hline
KOI & \# of tranets & $\langle P_j/P_{j-1} \rangle$ & ${\cal C}$ \\  
\hline  
   82       &         5  & 1.52 & 0.053 \\    
   111      &         3  & 2.13 & 0.004 \\  
   116      &         4  & 1.93 & 0.036 \\  
   117      &         4  & 1.67 & 0.038 \\  
   137      &         3  & 2.06 & 0.020 \\  
   168      &         3  & 1.47 & 0.020 \\  
   232      &         5  & 1.78 & 0.064 \\  
   271      &         3  & 1.85 & 0.093 \\  
   279      &         3  & 1.95 & 0.018 \\  
   282      &         3  & 3.22 & 0.001 \\  
   283      &         3  & 1.65 & 0.014 \\  
   408      &         4  & 2.10 & 0.070 \\  
   435      &         5  & 2.02 & 0.064 \\  
   474      &         4  & 2.70 & 0.049 \\  
   509      &         3  & 3.10 & 0.031 \\  
\hline \hline
\end{tabular}
}
\caption{KOIs with the observed multiplicities $m_{\rm obs} \geq 3$ and gap complexities ${\cal C}<0.1$. 
The third column reports the mean period ratio for planet pairs observed in each system,
$\langle P_j/P_{j-1} \rangle$. The full table 
in machine readable format is available for download.}
\end{table}

\begin{table}
\centering
{
\begin{tabular}{lcccc}
\hline \hline
    & Parameter        & Units      &  Value             &  Prior bounds          \\                      
\hline                    
(1) & $\lambda_{\rm pl}$  & --        & $3.05^{+0.15}_{-0.19}$       & $(1,7)$     \\ 
(2) & $f_{\rm H}$        & --         & $29.29^{+0.86}_{-0.77}$      & $(15,35)$  \\ 
(3) & $\sigma_{\rm H}$   & --         & $0.3864^{+0.044}_{-0.050}$    & $(0,1.5)$  \\           
(4) & $P_{\rm break}$     & days       & $6.48^{+2.57}_{-1.05}$       & $(3,15)$   \\           
(5) & $\alpha_P$        & --         & $-0.109^{+0.339}_{-0.337}$   & $(-1,1.5)$  \\  
(6) & $\beta_P$         & --         & $-0.889^{+0.023}_{-0.032}$   & $(-1.5,0)$   \\
\hline
(7) & $f_{\rm AMD}$       & --         & 1.0                     &  --        \\
(8) & $R_{\rm break}$     & $R_\oplus $  & $3.0$                    &  --        \\  
(9) & $\alpha_R$        & --         & $-1.5$                  & --         \\  
(10) & $\beta_R$         & --         & $-8.0$                  & --         \\
(11) & $\sigma_R$       & --         & $0.22$                  & --         \\
(12) & $\nu_R$          & --         & $0.5$                   & --         \\  
(13) & $\sigma_e$       & --          & $0.25$                  & --         \\ 
\hline \hline
\end{tabular}
}
\caption{Model ${\cal M}_{220}$ with the Poisson multiplicity distribution and $f_{\rm AMD}=0.3$. 
The prior bounds of parameters (1) to (6) are given in the last column. Parameters (7) to (13) 
were fixed in ${\cal M}_{220}$ to the values given in the fourth column. }
\end{table}

\begin{table}
\centering
{
\begin{tabular}{lcccc}
\hline \hline
    & Parameter        & Units      &  Value             &  Prior bounds          \\                      
\hline                    
(1) & $\beta_{\rm zipf}$  & --        & $-0.760^{+0.072}_{-0.072}$       & $(-2,0)$     \\ 
(2) & $f_{\rm H}$         & --        & $27.68^{+0.83}_{-0.81}$      & $(15,35)$  \\ 
(3) & $\sigma_{\rm H}$   & --         & $0.3433^{+0.041}_{-0.039}$    & $(0,1.5)$  \\           
(4) & $P_{\rm break}$     & days       & $6.11^{+1.35}_{-0.82}$       & $(3,15)$   \\           
(5) & $\alpha_P$        & --         & $-0.053^{+0.333}_{-0.230}$   & $(-1,1.5)$  \\  
(6) & $\beta_P$         & --         & $-0.868^{+0.027}_{-0.030}$   & $(-1.5,0)$   \\
\hline
(7) & $f_{\rm AMD}$       & --         & 1.0                     &  --        \\
(8) & $R_{\rm break}$     & $R_\oplus $  & $3.0$                    &  --        \\  
(9) & $\alpha_R$        & --         & $-1.5$                  & --         \\  
(10) & $\beta_R$         & --         & $-8.0$                  & --         \\
(11) & $\sigma_R$       & --         & $0.22$                  & --         \\
(12) & $\nu_R$          & --         & $0.5$                   & --         \\  
(13) & $\sigma_e$       & --          & $0.25$                  & --         \\ 
\hline \hline
\end{tabular}
}
\caption{Model ${\cal M}_{221}$ with the Zipfian multiplicity distribution and $f_{\rm AMD}=1$. 
The prior bounds of parameters (1) to (6) are given in the last column. Parameters (7) to (13) 
were fixed in ${\cal M}_{221}$ to the values given in the fourth column. }
\end{table}

\clearpage 
\begin{table}
\centering
{
\begin{tabular}{lcccc}
\hline \hline
    & Parameter        & Units      &  Value             &  Prior bounds          \\                      
\hline                    
(1) & $\beta_{\rm zipf}$  & --        & $-0.926^{+0.068}_{-0.077}$       & $(-2,0)$     \\ 
(2) & $f_{\rm H}$         & --        & $30.01^{+0.94}_{-0.87}$      & $(15,35)$  \\ 
(3) & $\sigma_{\rm H}$   & --         & $0.3739^{+0.044}_{-0.043}$    & $(0,1.5)$  \\           
(4) & $P_{\rm break}$     & days       & $10.36^{+2.75}_{-2.27}$       & $(3,15)$   \\           
(5) & $\alpha_P$        & --         & $-0.295^{+0.269}_{-0.132}$   & $(-1,1.5)$  \\  
(6) & $\beta_P$         & --         & $-0.987^{+0.050}_{-0.052}$   & $(-1.5,0)$   \\
\hline
(7) & $f_{\rm AMD}$       & --         & 1.0                     &  --        \\
(8) & $R_{\rm break}$     & $R_\oplus $  & $3.0$                    &  --        \\  
(9) & $\alpha_R$        & --         & $-1.5$                  & --         \\  
(10) & $\beta_R$         & --         & $-8.0$                  & --         \\
(11) & $\sigma_R$       & --         & $0.22$                  & --         \\
(12) & $\nu_R$          & --         & $0.5$                   & --         \\  
(13) & $\sigma_e$       & --          & $0.25$                  & --         \\ 
\hline \hline
\end{tabular}
}
\caption{Model ${\cal M}_{222}$ with the Zipfian multiplicity distribution and $f_{\rm AMD}=0.3$. 
The prior bounds of parameters (1) to (6) are given in the last column. Parameters (7) to (13) 
were fixed in ${\cal M}_{222}$ to the values given in the fourth column. }
\end{table}

\begin{table}
\centering
{
\begin{tabular}{lcccc}
\hline \hline
    & Parameter        & Units      &  Value             &  Prior bounds          \\                      
\hline                    
(1) & $\lambda_{\rm pl}$  & --        & $3.25^{+0.14}_{-0.14}$       & $(1,7)$     \\ 
(2) & $f_{\rm H}$        & --         & $30.9^{+1.1}_{-1.1}$        & $(15,35)$  \\ 
(3) & $\sigma_{\rm H}$   & --         & $0.3432^{+0.053}_{-0.046}$   & $(0,1.5)$  \\           
(4) & $P_{\rm break}$     & days       & $5.97^{+0.97}_{-0.68}$       & $(3,15)$   \\           
(5) & $\alpha_P$        & --         & $-0.004^{+0.341}_{-0.276}$   & $(-1,1.5)$  \\  
(6) & $\beta_P$         & --         & $-0.876^{+0.031}_{-0.033}$   & $(-1.5,0)$   \\
(7) & $f_{\rm AMD}$       & --         & $0.623^{+0.119}_{-0.099}$    & $(0,1)$     \\
\hline
(8) & $R_{\rm break}$     & $R_\oplus $  & $3.0$                    &  --        \\  
(9) & $\alpha_R$        & --         & $-1.5$                  & --         \\  
(10) & $\beta_R$         & --         & $-8.0$                  & --         \\
(11) & $\sigma_R$       & --         & $0.22$                  & --         \\
(12) & $\nu_R$          & --         & $0.5$                   & --         \\  
(13) & $\sigma_e$       & --          & $0.25$                  & --         \\ 
\hline \hline
\end{tabular}
}
\caption{Auxiliary model ${\cal M}_{228}$ with the Poisson multiplicity distribution. 
The prior bounds of parameters (1) to (7) are given in the last column. Parameters (8) to (13) 
were fixed in ${\cal M}_{228}$ to the values given in the fourth column. }
\end{table}

\clearpage

\clearpage

\clearpage

\clearpage

\clearpage

\clearpage

\clearpage

\clearpage

\clearpage

\clearpage

\clearpage

\clearpage

\clearpage

\clearpage

\end{document}